\documentclass[aps, pra, twocolumn,10pt,showkeys]{revtex4-1} 	
\pdfoutput=1
\usepackage[unicode]{hyperref}
\usepackage{dcolumn}
\usepackage{bm}
\usepackage{graphicx}
\usepackage{color}
\usepackage{amsmath}
\usepackage[T2A]{fontenc}
\usepackage[utf8x]{inputenc}
\usepackage[english]{babel}
\usepackage{longtable}
\usepackage{float}
\usepackage{soul}
\usepackage{multirow}
\usepackage{xcolor}
\usepackage{xfrac}

\begin{document}

\title{Inner-shell magnetic dipole transition in Tm atom\\ as a candidate for optical lattice clocks}

\author{D.\,Sukachev,$^{1,\,2,\,3}$ S.\,Fedorov,$^{4}$ I.\,Tolstikhina,$^{1,\,5}$ E.\,Kalganova,$^{1,\,2,\,5}$ 
G.\,Vishnyakova,$^{1,\,2,\,5}$ K.\,Khabarova,$^{1,\,2}$ D.\,Tregubov,$^{1,\,5}$ A.\,Golovizin,$^{1,\,2,\,5}$ V.\,Sorokin,$^{1,\,2}$ 
N.\,Kolachevsky$^{1,\,2,\,5}$}
\email{kolachbox@mail.ru}

\affiliation{
$^1$\,P.N.\,Lebedev Physical Institute, Leninsky prospekt 53, 119991 Moscow, Russia\\
$^2$\,Russian Quantum Center,  Business-center ``Ural'', 100A, Novaya street, Skolkovo, Moscow, 143025 Russia \\
$^3$\,Harvard University, Physics Department, 17 Oxford str., Cambridge MA, 02138 USA\\
$^4$\,Ecole polytechnique fédérale de Lausanne, Route Cantonale, 1015 Lausanne, Switzerland\\
$^5$\,Moscow Institute of Physics and Technology, 141700 Dolgoprudny, Moscow region, Russia\\
}

\date{today}

\begin{abstract}
We consider a narrow magneto-dipole transition in the $^{169}$Tm atom at the wavelength of 1.14\,$\mu$m as a candidate for a 2D  
optical lattice clock. Calculating dynamic polarizabilities of the two clock levels $[\text{Xe}]4f^{13}6s^2 (J=7/2)$ and  $[\text{Xe}]4f^{13}6s^2 
(J=5/2)$  in the spectral range from 250\,nm to 1200\,nm, we suggest the ``magic'' wavelength for the optical lattice at  807\,nm.
Frequency shifts due to  black-body radiation (BBR), the van der Waals interaction, the magnetic dipole-dipole interaction, and other 
effects which can perturb the transition frequency are calculated. The transition at 1.14\,$\mu$m demonstrates low sensitivity to 
the BBR shift corresponding to  $8\times10^{-17}$ in fractional units at  room temperature which makes it an
interesting candidate for high-performance optical clocks. The total estimated frequency uncertainty is less than $5 \times 10^{-18}$ in fractional units. By direct excitation of the 1.14\,$\mu$m transition in Tm atoms loaded 
into an optical dipole trap, we set the lower limit for the lifetime  of the upper clock level $[\text{Xe}]4f^{13}6s^2 (J=5/2)$ of 
$112$\,ms which corresponds to a natural spectral linewidth narrower  than 1.4\,Hz. The polarizability of the Tm ground state was 
measured by the excitation of parametric resonances in the optical dipole trap at 532\,nm.
\end{abstract}



\maketitle

\tableofcontents

\section{Introduction}

Magnetic-dipole transitions between the ground state fine structure components in hollow shell lanthanides are strongly shielded from 
external electric fields by the closed outer $5s^2$ and $6s^2$ shells. In the solid state these well resolved transitions protected from 
intra-crystal electric fields are widely used in various active media doped by  Er$^{3+}$, Tm$^{3+}$, and other ions lasing in the 
near-infrared and infrared spectral ranges~\cite{Zharkov1975, Barnes1993}. Such shielding can also facilitate the use of inner-shell 
transitions in optical frequency metrology due to low sensitivity to external electric fields and collisions~\cite{Boettger2001}.

In 1983 Alexandrov et. al.~\cite{ref:AlexandrovEnglish1983} showed that  the collisional broadening of the inner-shell magnetic dipole 
transition in the Tm atom $[\text{Xe}]4f^{13}6s^2 (J=7/2) \rightarrow [\text{Xe}]4f^{13}6s^2 (J=5/2)$, where $J$ is the total electronic angular momentum, at the wavelength of 1.14\,$\mu$m in He buffer 
gas is  suppressed by at least  500 times compared to the outer shell transitions. Note, that in the early era of optical atomic 
clocks the dominating systematic uncertainty was the collisional shift in a cloud of laser cooled atoms~\cite{Wilpers2002, Ido2005}.
One could expect better performance using  inner-shell transitions in lanthanides, but this study was hampered by difficulties with 
their laser cooling. It was shown later that  for Tm-He collisions shielding strongly reduces the spin relaxation~\cite{Hancox2004} 
but it does not reduce the spin relaxation rate in Tm-Tm collisions due to the anisotropic nature of the magnetic dipole-dipole 
interaction~\cite{Connolly2010}.

The problem with atom-atom collisions in optical clocks was solved after the invention of an optical lattice clock~\cite{Katori2003, 
Takamoto2003} which resulted in rapid progress of accuracy and stability over the last decade~\cite{Ludlow2015}. Today, lattice clocks 
based on Sr~\cite{Bloom2014} and Yb~\cite{Hinkley2013} demonstrate unprecedented fractional frequency instabilities in the low 
$10^{-18}$ range. One of the important limiting factors is the shift caused by black-body radiation (BBR)~\cite{Beloy2014, 
Safronova2013, Ushijima2015}.

Optical clock community continues an intensive search for alternative  candidates aiming for lower sensitivity to BBR and other shifts, 
simplicity of manipulation and better accuracy~\cite{Kulosa2015, McFerran2014}. Since hollow-shell lanthanides are expected to show 
small differential static polarizabilities of the states with different configurations of the $4f$ electrons, one expects small BBR 
shift of the inner-shell magnetic dipole transitions. Taking into account large natural lifetimes of the clock levels, these 
transitions can be successfully used in  optical lattice clocks. Recent progress in laser cooling of Er~\cite{Aikawa2012}, 
Dy~\cite{ref:DyBEC:Lev:PRL}, and Tm~\cite{Sukachev2010, Sukachev2014} and frequency stabilized laser systems~\cite{Alnis2008, 
Kessler2012} open the way for experimental implementation of these ideas.

\begin{figure}
\includegraphics[width=1\linewidth]{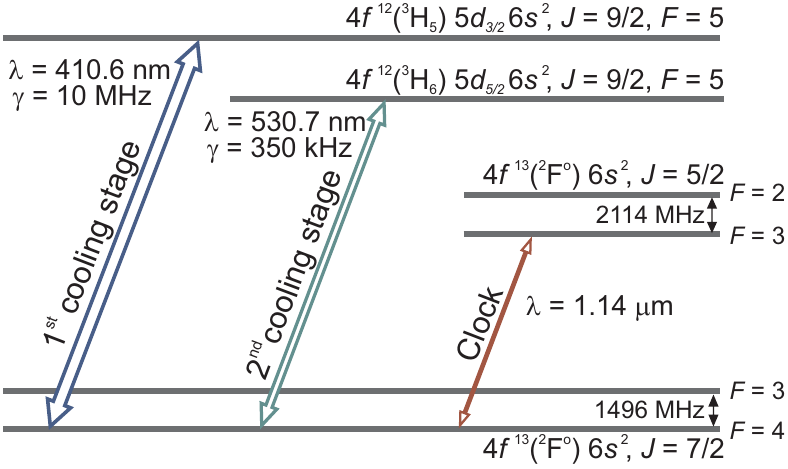}
\caption{\label{fig_levels} Relevant energy levels of $^{169}$Tm. The strong transition at 410.6\,nm is used for the first-stage laser 
cooling and detecting the ground state populations and the weak  transition at 530.7\,nm is used for the second-stage cooling. The 
proposed clock transition $4f^{13}6s^2 (J=7/2, F=4) \rightarrow 4f^{13}6s^2 (J=5/2, F=3)$ is at the wavelength of 1.14\,um.}
\end{figure}

Similar to other lanthanides, laser cooling of Tm is achieved in two stages. 
The first cooling stage is done at the strong 410.6\,nm 
transition which routinely allows reaching   subdoppler temperature of 80\,$\mu$K in a cloud of $2\times10^6$ atoms~\cite{Sukachev2010}. 
The second cooling stage at the weak 530.7\,nm transition  results in the Doppler-limited temperature of 9\,$\mu$K~\cite{Vishnyakova2016}. 
This temperature is low enough to load atoms in a shallow optical trap or a lattice as was demonstrated in \cite{Sukachev2014} using 
532\,nm laser radiation. 
Relevant Tm levels are shown in Fig.~\ref{fig_levels}. 
Further cooling of atoms is possible by the optimization of the cooling sequence~\cite{Frish-PhD-2014} or by evaporative cooling~\cite{Ketterle1996}. These experiments stimulated further study of the inner shell transition $[\text{Xe}]4f^{13}6s^2 (J=7/2, \,F=4) \rightarrow [\text{Xe}]4f^{13}6s^2 (J=5/2, \,F=3)$, where $F$ is the total atomic angular momentum, for its application in optical lattice clocks. 
In this article, the level $[\text{Xe}]4f^{13}6s^2 (J=7/2)$ will be referred to as the ``lower clock level'' while the level $[\text{Xe}]4f^{13}6s^2 (J=5/2)$ as the ``upper clock level''.

In the next sections, we analyze effects which may impact the performance of such clocks. 
First, the only stable isotope $^{169}$Tm is a boson and the clock transition is subject to collisional shifts. The related scattering length depends on the poorly known Tm-Tm potential at small distances and is very sensitive to the calculation uncertainty of the long-range potentials~\cite{Dalibard1998,Gribakin1993}. This difficulty can be overcome if Tm atoms are loaded in a 2D-optical lattice with a small filling factor canceling Tm-Tm collisions.
Second, to avoid intensity-dependent effects we calculated the dynamic polarizabilities of the upper and lower clock levels and defined a candidate  for the ``magic'' wavelength (sec.~\ref{sec:polarizabilities}). 
Third, the large ground-state dipole moment of Tm atoms induces a frequency shift due to 
magnetic dipole-dipole interaction. Preparing Tm atoms in the $|m = 0\rangle$ (here $m$ is a magnetic quantum number) state cancels this shift but magnetic relaxation still limits the interrogation time of the clock transition and should be 
taken into account (sec.~\ref{sec:magnetic}). 
In sec.~\ref{sec:budget} we present the error budget of the proposed Tm optical clock.

In the experimental part (sec.~\ref{sec:experiment}), we demonstrate direct excitation of the clock transition at 1.14\,$\mu$m  and measure the lifetime of the upper clock level in a 1D optical lattice formed by 532\,nm laser 
radiation~\cite{Golovizin2015}. Also, we experimentally evaluate the dynamic polarizability  of the Tm ground state at 532\,nm by 
excitation of  parametric resonances in the optical dipole trap.

\section{Polarizabilities}
\label{sec:polarizabilities}
\label{sec:polarizability:theory}
To find the magic wavelength and to estimate the BBR and the van der Waals shifts, one should know the energy shifts $\varDelta E$ of the 
clock states in an external monochromatic electric field $\vec{E} = \sfrac{1}{2} \vec{\mathcal{E}} e^{-i\omega 
t}+\text{c.c.} $ at the angular frequency $\omega$

\begin{equation}
		\varDelta E(\omega) = -\frac{\alpha(\omega)}{4} |\mathcal{E}|^2-\frac{\gamma(\omega)}{64} |\mathcal{E}|^4 +...,
		\label{eq1}
	\end{equation}
where $\alpha(\omega)$ is the dynamic polarizability  and $\gamma(\omega)$ is the hyperpolarizability, both depending on $m$ 
and the polarization of the field.  To our knowledge, there are only a few publications where the polarizability of Tm levels was 
analyzed. In \cite{RETensorPol, TmTensorPol} the authors measured the  static tensor polarizability, while a theoretical calculation of 
static polarizabilities without accounting for a fine-structure interaction is presented in~\cite{Chu2007}. In this section we will 
calculate polarizabilities of the clock states.

\begin{figure}
\includegraphics[width=1\linewidth]{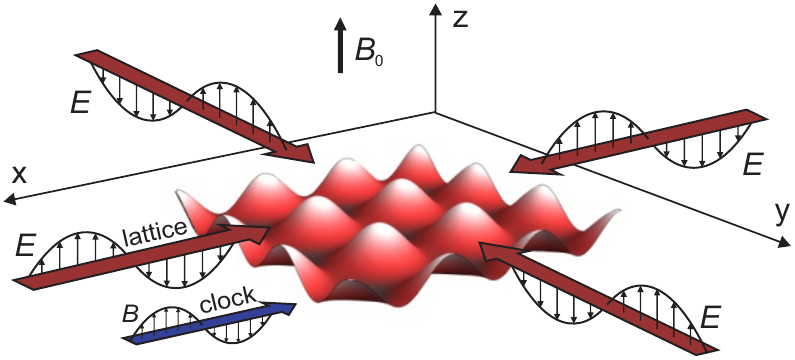}
\caption{\label{latticeGeometry}
Geometry of a 2D optical lattice. The lattice is formed by 4 horizontal laser beams (lattice, red) with linear vertical 
polarization. A uniform external magnetic field $B_0$ is applied along the vertical axis. An interrogating laser beam  (clock, blue) 
lies in the optical lattice plane (horizontal) to eliminate frequency shifts by the Doppler-effect and photon recoil. The  $B$-component of the 
interrogating light should be  vertical   to excite the $|m=0 \rangle \rightarrow |m=0\rangle$ clock transition (the clock transition is of a magnetic dipole type). 
}
\end{figure}

To suppress the site-dependent frequency shift from varying light polarization in the lattice, we suggest loading Tm atoms into a 2D 
optical lattice formed by 4 laser beams with the same linear polarization as shown in Fig.\,\ref{latticeGeometry}. This guarantees that the trapping light polarization is the same for all lattice sites. Further in this paper, we consider only the transition 
$|J=7/2,F=4, m=0\rangle \rightarrow |J=5/2,F=3,m=0\rangle$  which is free from the  frequency shifts induced by the magnetic dipole-dipole interaction (see sec.\,\ref{sec:magnetic}). Since both levels have $m=0$, the contribution from the vector polarizability for this transition also vanishes  and the total 
polarizability $\alpha$ can be separated into the scalar $\alpha^s$ and the tensor $\alpha^t$ parts~\cite{Lepers2014} as follows:
\begin{equation}
\begin{split}
	\alpha_{JFm}(\omega) &= \alpha^s_{JF}(\omega) + \alpha^t_{JF}(\omega) \frac{3m^2-F(F+1)}{F(2F-1)}\,,\\
	\alpha^s_{JF}(\omega) &=\frac{1}{2F+1}\sum\limits_{m=-F}^{m=F} \alpha_{JFm}(\omega)\,,\\
	\alpha^t_{JF}(\omega) &= \alpha_{JF,m=F}(\omega)-\alpha^s_{JF}(\omega).
\end{split}
	\label{eq2}
\end{equation}
For consistency with other papers, we will calculate the polarizabilities in atomic units (a.u.); 1\,a.u. = 
$4\pi \epsilon_0 a_0^3$	= $1.65\times 10^{-41}$\,J/(V/m)$^2$,
where $a_0$ is the Bohr radius and $\epsilon_0$ is the vacuum permittivity (for conversion to another units, see~\cite{Mitroy2010}).

\subsection{Discrete spectrum}

The contribution of a discrete spectrum is given by~\cite{Lepers2014, Angel1968}
\begin{multline}\label{eq:alphaFM}
\alpha_{Fm}(\omega)=\frac{3}{2} \frac{c^3\hbar^4}{a_0^3} \sideset{}{'} \sum_{F'}\frac{2F_u+1}{(E_{F'}-E_F)^2}
	\left(
		\begin{matrix}
			F_u & 1 & F_d\\
			-m  & 0  & m
		\end{matrix}
	\right)^2\\
 \times \frac{A_{F_u \to F_d}}{(E_{F'}-E_F)^2-(\hbar\omega)^2},
\end{multline}
where $c$ is the speed of light, $\hbar$ is the reduced Planck’s constant, and $E_F$ and $E_{F'}$ are energies of levels $|F\rangle$ 
and $|F'\rangle$, respectively.   The summation is over all levels $F'$. For each term, $F_u = F'$ and $F_d = F$ if $E_{F'} > E_{F}$ 
and vice versa. $A_{F_u \to F_d}$ is a transition probability (spontaneous decay rate) from $|F_u\rangle$ to $|F_d\rangle$.

Assuming $JI$-coupling between the total electron momentum $J$ and the nuclear spin $I$, the scalar polarizability is independent of 
$F$~\cite{Lepers2014, Angel1968}:
\begin{multline}
	\alpha^s_{JF}(\omega)=\alpha^s_{J}(\omega)=\frac{1}{2J+1}\sum\limits_{m_J=-J}^{m_J=J} \alpha_{Jm_J}(\omega)=\\
	\frac{1}{2} \frac{c^3}{a_0^3} \sideset{}{'}\sum_{J'}\frac{2J_u+1}{2J_d+1}\frac{1}{(\omega_{J'J})^2}  \frac{A_{J_u \to 
J_d}}{(\omega_{J'J})^2-\omega^2},
\end{multline}
where $\omega_{J'J}=(E_{J'}-E_J)/\hbar$.
The tensor polarizability equals
\begin{multline}\label{eq:alphaJFt}
	\alpha^t_{JF}(\omega) = \alpha^t_{J}(\omega) \times (-1)^{I+J+F} 	\left\{
		\begin{matrix}
			F & J & I\\
			J  & F  & 2
		\end{matrix}
	\right\} \\
	\times \sqrt{\frac{F(2F-1)(2F+1)(2J+3)(2J+1)(J+1)}{(2F+3)(F+1)(2J-1)J}},
\end{multline}
where
\begin{multline}
	\alpha^t_{J}(\omega) = \frac{3c^3}{a_0^3} \sideset{}{'} \sum\limits_{J'} \frac{2J_u+1}{\omega_{J'J}^2} \frac{A_{J_u\rightarrow 
J_d}}{\omega_{J'J}^2-\omega^2} 	 (-1)^{J+J'} \\
		\times \left\{
		\begin{matrix}
			1 & 1 & 2\\
			J_d  & J_d  & J_u
		\end{matrix}
	\right\} \sqrt{\frac{5J(2J-1)}{6(J+1)(2J+1)(2J+3)}}.
\end{multline}
Note that 
$\alpha^t_{\sfrac{7}{2},\,4} = \alpha^t_{\sfrac{7}{2}}$ and $\alpha^t_{\sfrac{5}{2},\,3} = \alpha^t_{\sfrac{5}{2}}$.

 As an input for calculation of $\alpha$ one should have  transition probabilities  $A_{J'\rightarrow J}$ from the level of interest 
 to all others.
Though many transition wavelengths in the spectral range  from 250\,nm to 807\,nm and their probabilities were measured 
in~\cite{TmTransitions, Anderson1996} by Fourier-transform spectroscopy and time-resolved laser-induced fluorescence, there is  still 
a number of  transitions in the UV, visible, and IR spectral ranges which are essential for calculation and have unknown probabilities. We used the 
numerical package COWAN~\cite{Cowan1981} to calculate transition wavelengths and probabilities in the spectral range from 250\,nm to 
1200\,nm  (see Appendix A).

The most self-consistent approach for calculation of the  differential static 
polarizability of the clock levels is to use only the numerically calculated wavelengths and probabilities.
A slight modification of this approach is used for calculation of the  magic wavelengths 
(sec.~\ref{sec:mw}).

As expected from general considerations concerning the inner shell transitions, the static scalar polarizabilities for the  clock levels are nearly equal. Our calculation shows that they differ by less than $0.1$\,a.u. and are equal to 138\,a.u. 
Note, that the calculated static tensor polarizability of  $-2.7$\,a.u. for the lower $|J=7/2\rangle$ clock level  is in  good agreement with 
the  known  experimental value of $-2.7(2)$\,a.u.~\cite{RETensorPol}. For the upper $|J=5/2\rangle$ clock level our calculations give $-2.3$ atomic units for the tensor static polarizability.

\subsection{Continuous spectrum}
\begin{figure}
\includegraphics[width=1\linewidth]{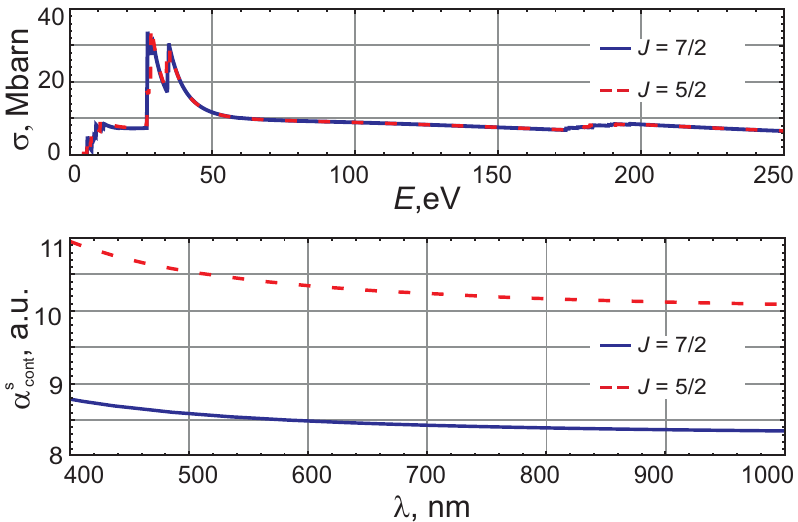} 
\caption{\label{fig_ion} Upper panel: The photoionization cross sections for  $|J=7/2\rangle$ (solid blue) and  $|J=5/2\rangle$ (dashed red)  clock levels. Peaks 
around 28 and 35\,eV correspond to strong resonance enhancement~\cite{ref:Whitfield2008}.
 Lower panel: Contribution to the scalar polarizability $\alpha^s_\text{cont}(\omega)$ from the continuous spectra. The hyperfine interaction is not taken into account.}
\end{figure}

 To determine a contribution of the continuous spectrum (ignoring hyperfine interaction) to the polarizability  we used the formula~\cite{Veseth1992}:
\begin{equation}
\alpha^s_{\text{cont}}(\omega) = \frac{c}{2\pi^2}\int^{\infty}_{\omega_I} 
\frac{\sigma(\omega')d\omega'}{(\omega'-\omega_n)^2-\omega^2},
\end{equation}
where $\omega_I$ is the photoionization limit and $\sigma(\omega)$ is the photoionization cross section of the energy level.
The ionization cross-section was numerically calculated  using the package FAC ~\cite{FAC2011} and the results  are shown in the upper panel of
Fig.\,\ref{fig_ion}.  Using these results, we evaluated  the polarizabilities $\alpha^s_\text{cont}(\omega)$ for the clock levels resulting from transitions to continuous spectrum (Fig.\,\ref{fig_ion}, lower panel).

The contributions  $\alpha^s_\text{cont}(\omega)$  are  small compared to the contribution from the discrete spectrum and differ only 
by 2\,a.u. for the two fine structure components $|J=7/2\rangle$ and $|J=5/2\rangle$. This means that the transitions to continuum  basically do not influence positions of  magic wavelengths. We also assume that the corresponding contribution to the tensor polarizability is even 
smaller and will neglect it in further analysis. Since the discrete spectrum gives the equal static scalar polarizabilities for the clock levels, we expect the continuous spectrum to contribute to polarizabilities of the both levels equally as well. 
Thus, a rough estimation of the error of calculated contribution of the continuous spectrum to the differential polarizability is about the difference between $\alpha^s_\text{cont}(0)$ for the clock levels, i.e., 2\,a.u.
Unfortunately, we do not know of any experimental data on the photoionization 
cross sections for Tm atoms and therefore can't rigorously estimate the error.

\subsection{Magic wavelength}
\label{sec:mw}

The magic wavelengths for optical traps providing the vanishing  total light shift of the clock transition (\ref{eq1}) are widely used in 
optical clocks~\cite{Katori2003}. To determine the magic wavelengths  one should search  for the crossing points of the dynamic 
polarizabilities for the upper and lower clock levels (neglecting the  contribution from  the hyperpolarizability in the first 
approximation). Position of the magic wavelengths strongly depend on energies and probabilities of the resonances in the atom. In 
general, we can not use  only results of our calculations because of insufficient accuracy provided by the COWAN package (see Appendix A).

 To solve  this problem first we tried a ``combined'' approach. Calculated transitions were assigned with experimental ones which can 
 be done  unambiguously for  wavelengths $\lambda>500$\,nm. It turns to be impossible  for the shorter wavelengths  ($\lambda<500$\,nm) due to 
 higher density of transitions. Then we combined the calculated spectrum for $\lambda<500$\,nm, the available experimental data for 
 $\lambda>500$\,nm, and calculated probabilities for known transitions but with not measured probabilities. After detailed study we 
 concluded that it is a questionable approach because the calculated and experimentally measured transition probabilities sometimes 
 differ by an order of magnitude (see Fig.\,\ref{fig:COWAN}, lower panel). This difference
impacts the calculated polarizabilities in a wide spectral range impeding reliable prediction of the magic wavelengths.

  To our opinion, a more reliable approach bases on maximal use of calculation results:  We took the calculated spectrum and substituted 
  the predicted wavelengths with correct ones known from the experiment for all transitions with $\lambda>500$\,nm. As for the 
  probabilities, we used the calculated ones except the case when the probability is smaller than $10^5$\,s$^{-1}$. This method give 
  reliable results for the  magic wavelengths in the near IR region with low density of strong transitions.

\begin{figure}
\includegraphics[width=1\linewidth]{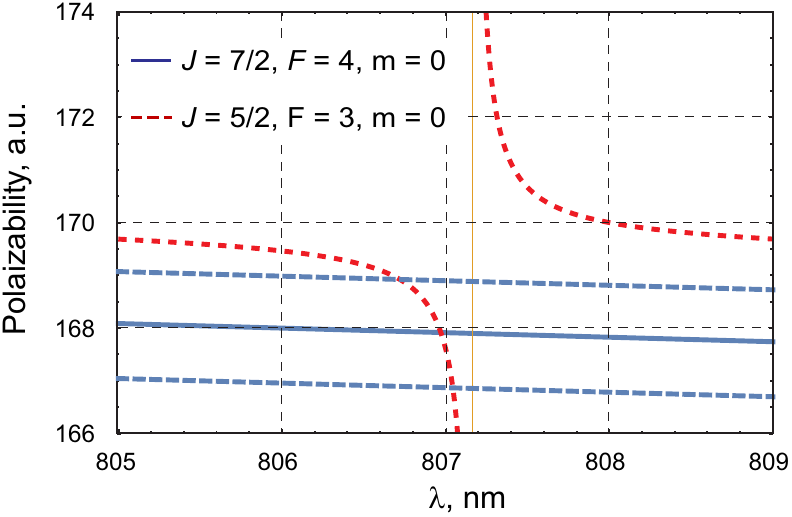}
\caption{The magic wavelengths for the 1.14\,$\mu$m clock transition in Tm atom ($m=0\rightarrow m=0$) around 807\,nm 
calculated for the  linear vertical polarization of the trapping light (see Fig.\,\ref{latticeGeometry}). Blue solid curve is the polarizability of the lower 
$|J=7/2,F=4,m=0\rangle$ clock state, red dashed curve is the polarizability of the upper $|J=5/2,F=3,m=0\rangle$ clock state. The dashed 
blue lines show  the anticipated uncertainty in the calculation of the  differential polarizability which may impact the  position of 
the magic wavelength.
}
\label{fig:someMW}
\end{figure}

Selected approach predicts the reliable
candidate for the magic wavelength at 807\,nm with an attractive lattice potential (Fig.~\ref{fig:someMW}). 
Its  presence is caused by 
the weak transition from the $|J=5/2\rangle$ clock level $ 4f^{13}(^2F^o)6s^2 (J=5/2)$ with the energy  $8771.24\,\text{cm}^{-1}$ to another 
level $4f^{12}(^3F_4)5d_{3/2}6s^2  (J=5/2)$, with the energy $21161.4\,\text{cm}^{-1}$ at 807.1\,nm ~\cite{TmTransitions}. At the same 
time, there are no allowed transition from the $|J=7/2\rangle$  clock level in the vicinity of 807\,nm. 
Taking into account the uncertainty in the contribution of the continuous spectra to the evaluated differential polarizability of $\pm 
1$\,a.u., the proposed magic wavelength should be blue-detuned from the  transition 807.1\,nm  by  0.1\,nm --- 1\,nm.

The  figure-of-merit for an optical lattice comes from its  depth, the off-resonant scattering rate and the magnetic dipole-dipole 
relaxation rate.
The  optical lattice depth in kelvin is given by:
\begin{equation}
U\,[\text{K}] = \alpha[\text{a.u.}]\frac{2\pi a_0^3}{c\,k_B}I\,[\text{W/m$^2$}],
\end{equation}
where $I$ is the field intensity in lattice anti-nodes given in W/m$^2$ and  $k_B$ is the Boltzmann constant.
The spontaneous decay following the off-resonant excitation by the lattice field perturbs the coherence of the clock levels and should be taken into account. The off-resonance scattering rate for the transition $|m=0\rangle \rightarrow |m=0\rangle$ can be estimated as~\cite{Lepers2014}:
\begin{multline}
\varGamma(\omega)_{0\rightarrow 0} = I \sideset{}{'} \sum \limits_{F'} \frac{\omega_{F'F}^2+\omega^2}{\left[\omega_{F'F}^2-\omega^2 \right]^2}
 \frac{3\pi c^2\,A_{F' \rightarrow F}}{\hbar \omega_{F'F}^3} \\
\times
\left(
		\begin{matrix}
			F_u & 1 & F_d\\
			0 & 0  & 0
		\end{matrix}
	\right)^2 (2F_u+1) \varGamma_{F_u},
\label{eq:alpha-scal-im}
\end{multline}
where $\varGamma_{F_u}$ is the inverse lifetime of  $|F_u \rangle$ level.

The optical lattice at 807\,nm can be formed by a Ti:sapphire laser beam.
With 0.5\,W output power focused in the beam 
waist of 50\,$\mu$m (radius at $1/e^2$ intensity level) corresponding to $I=50$\,kW/cm$^2$ in the retro-reflected configuration, one expects the trap depth of 20\,$\mu$K. 
This is  enough to capture Tm atoms from a narrow-line MOT. Even for the smallest expected detuning from the 807.1\,nm resonance of 
0.1\,nm, the off-resonant scattering rate is less than $0.1$\,s$^{-1}$.

\subsection{Hyperpolarizability}
\label{sec:hyper}
The magic wavelength depends not only on the differential polarizability of the clock states, but also on the differential 
hyperpolarizability (\ref{eq1}) and light intensity $I$. The scalar hyperpolarizability $\gamma(\omega)$ is given by~\cite{Bishop1994} :
\begin{equation}
\label{eq:hyper1}
\gamma(\omega) = \frac{1}{4} \left( \gamma^+(\omega) + \gamma^-(\omega) \right),
\end{equation}
where
\begin{multline}
\gamma^{+}=\frac{4}{\hbar^3} \sideset{}{'} \sum_{m,k,n}(D_z)_{gm}(D_z)_{mk}(D_z)_{kn}(D_z)_{ng}\\
\times\Bigg(\frac{4\omega_{mg}\omega_{ng}}{\omega_{kg}\left(\omega^2-\omega_{mg}^2\right)\left(\omega^2-\omega_{ng}^2\right)}\\
+\frac{1}{(\omega_{mg}-\omega)(\omega_{kg}-2\omega)(\omega_{ng}-\omega)}\\
+\frac{1}{(\omega_{mg}+\omega)(\omega_{kg}+2\omega)(\omega_{ng}+\omega)}\Bigg)
\label{eqXp2}
\end{multline}
and
\begin{multline}
\gamma^{-}=\frac{8}{\hbar^3} \sideset{}{'} \sum_{m,n}\left|(D_z)_{mg}\right|^2\left|(D_z)_{ng}\right|^2 \\ 
\times \frac{\omega_{mg}\left(\omega^2+3\omega_{ng}^2\right)}{\left(\omega^2-\omega_{mg}^2\right)\left(\omega^2-\omega_{ng}^2\right)^2}\,,
\end{multline}
where $(D_z)_{i,j}$ is a matrix element of the $z$-projection of the dipole moment between levels $i$ and $j$.

For calculation of  hyperpolarizability  we used  the transition matrix elements, their signs, and the transition wavelengths obtained by  the COWAN package  for all transitions except the 807.1\,nm one. For this transition, we used the experimentally measured wavelength and probability; the sign of the transition matrix element was taken from the numerical calculations.
This exception is done to  improve  accuracy of the magic wavelength 
prediction.  

\begin{figure}
\includegraphics[width=1\linewidth]{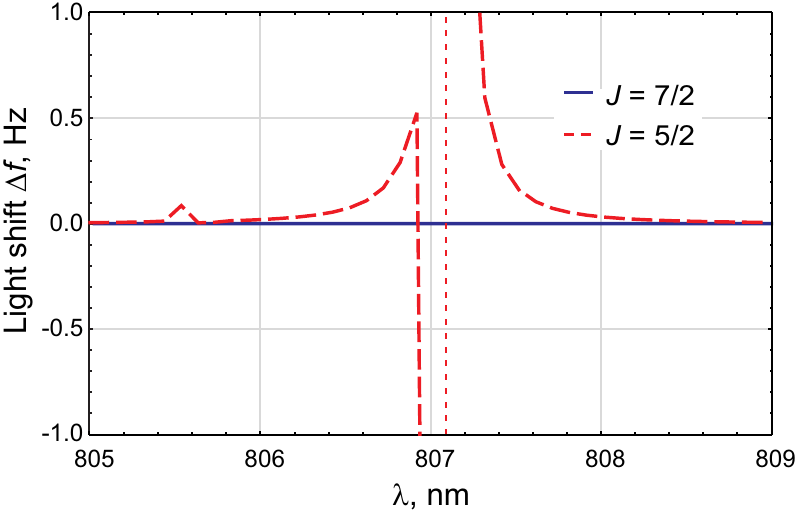}
\caption{Light shifts of the lower $|J=7/2\rangle$ (blue, solid) and upper $|J=5/2\rangle$ (red, dashed) clock levels caused by the  hyperpolarizability. Vertical red line denotes the position of the resonance.
Calculations are done for an intensity of $I=50$\,kW/cm$^2$; the hyperfine interaction is ignored.}
\label{fig:hyper}
\end{figure}

The light  shifts for the clock levels $|J=7/2\rangle$ and $|J=5/2 \rangle$ coming from  hyperpolarizability in the optical lattice at $\lambda= 807$\,nm and $I = 50$\,kW/cm$^2$ is shown in Fig.\,\ref{fig:hyper}.
As follows from the previous section,  the magic wavelength is blue detuned from the 807.1\,nm resonance by more than 
0.1\,nm, which makes the  hyperpolarizability shift to be less than $0.5$\,Hz. Corresponding correction to the magic wavelength is 
negligible. Still,  hyperpolarizability contributes to the clock frequency uncertainty which is discussed later in 
sec.\,\ref{sec:budget}.

\section{Magnetic interactions}
\label{sec:magnetic}
\subsection{Magnetic dipole-dipole interaction}
\label{sec:mdd}
The magnetic moment of the thulium ground state equals $4 \mu_B$ ($\mu_B$ is the Bohr magneton) which causes  a magnetic dipole-dipole 
interaction between atoms. The interaction potential between two atoms is
\begin{equation}
U_{dd}(r)=\mu_0 (g_F \mu_B)^2 \frac{\hat{\vec{F_1}} \cdot \hat{\vec{F_2}}-3(\hat{\vec{F_1}} \cdot \hat{\vec{r}}) (\hat{\vec{F_2}} 
\cdot \hat{\vec{r}})}{4\pi r^3},
\label{mdd}
\end{equation}
where $\vec{F}_{1,2}$ are the total atomic angular momenta, $\mu_0$ is the magnetic permeability of vacuum, $\vec{r}$ is the vector pointing 
from one atom to another and $g_F$ is the Land\'{e} g-factor of the ground state. For the Tm ground state  $g_F \approx 1$.

For spatially non-uniform 
atom distributions over optical lattice sites, the magnetic 
interaction may lead to inhomogeneous broadening and frequency shifts of the clock transition, both of them being of the same order of 
magnitude. These shifts correspond to the interaction energy between neighboring  atoms.
For two Tm atoms loaded in the adjacent sites of optical lattice at 800\,nm and prepared in the  $|m=4 \rangle$ magnetic state, the 
interaction energy (\ref{mdd}) corresponds to the frequency shift of
\begin{equation}
\label{eq:mdd_shift}
 \varDelta f_{dd} \approx \frac{\mu_0 (m\, \mu_B)^2 }{4\pi r^3 h} \,
 \end{equation}
which is of the order of 10\,Hz. The shift is large and difficult to predict due  to randomness of 
the lattice  site occupation. Further, we will analyze  only the transition $|m=0\rangle \rightarrow |m=0\rangle$ which is insensitive 
to this shift.

\begin{figure}
\includegraphics[width=1\linewidth]{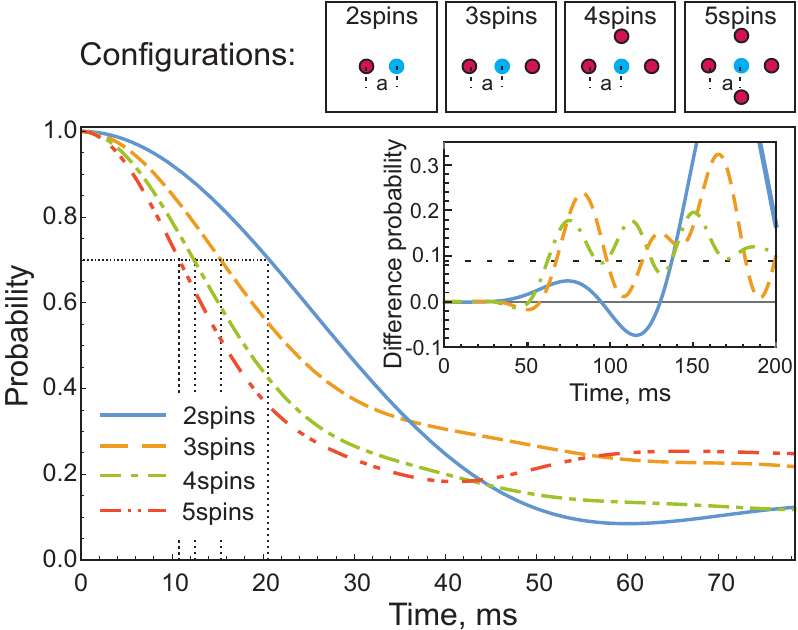}
\caption{Spin relaxation dynamics.
Upper panel: Configurations of atoms used in the simulation, $a=400$\,nm is the inter-atomic separation, quantization axes is 
perpendicular to the plane of the sketch;
Main panel: Probability to find the central atom (blue) in the state $|m=0 \rangle$ for 2, 3, 4, and 5 atoms in the 2D optical lattice. 
Note, that the probabilities do not go to zero at longer times because at steady state there is almost uniform distribution among 
magnetic sublevels.
Inset: Difference of the probabilities to find the first atom in the state $|m=0\rangle$ for exact and approximate solutions for 2,3, and 4 atoms.  The horizontal dashed line represents the difference of the steady-state probabilities (see the main text).
}
\label{fig_dipolar}
\end{figure}

Magnetic  dipole-dipole interactions also limit the interrogation time because of  spin relaxation: the atomic ensemble prepared 
into a pure polarized state  will gradually lose its polarization.
To evaluate the corresponding relaxation time, we solved the Shr\"{o}dinger equation with the interaction (\ref{mdd}) for 2, 3, 4, and 5 
spatially-fixed Tm atoms in ground state ($F_1 = F_2 = 4$) prepared in  the initial $|m=0\rangle \otimes ... \otimes |m=0\rangle $ state at the vanishing external  
magnetic field. The spatial separation of $a=400$\,nm corresponds to an 800-nm optical lattice. The relative positions of the atoms 
are shown in Fig.\,\ref{fig_dipolar}, upper panel. The lower panel of Fig.\,\ref{fig_dipolar} shows dynamics of the spin state for the central atom,  marked blue 
in the upper panel of Figure.

For 2, 3, and 4 atoms the Shr\"{o}dinger equation was solved exactly. For 5 atoms the Hilbert space is too large and we restricted our  
calculation to the subspace $m_i = \left\{-2,-1,0,1,2\right\}$ . To estimate validity of this approach we also solved the 
Shr\"{o}dinger equation for 2, 3, and 4 atoms in the restricted subspace. The inset in the lower panel of Fig.\,\ref{fig_dipolar} shows good agreement between 
approximate and exact solutions for the first 50\,ms of the evolution. In the steady-state, it is reasonable to assume that all spin projections are equiprobable. Consequently, average probability to find the central spin in the $|m=0 \rangle$ state  equals 1/9 for the full Hilbert space  and 1/5 for the truncated space.  This explains the discrepancy between the exact and approximate solutions at longer times (> 50\,ms). 
The characteristic relaxation time was derived by setting the probability to find the central spin in the initial $|m=0\rangle$ state 
to 0.7. It equals 20\,ms, 13\,ms, 11\,ms, and 10\,ms for 2,3,4, and 5 spins, respectively (Fig.\,\ref{fig_dipolar}).

External magnetic field reduces  spin relaxation because some spinflip processes require additional energy. A significant reduction of the spin 
relaxation is expected if the Zeeman splitting becomes  larger than the kinetic energy $E_K$ of atoms. At the experimentally achieved 
temperature of $T\sim 10$\,$\mu$K, the  kinetic energy equals  $E_K=k_B T \sim 100$\,kHz$\times h$. This energy  corresponds to a  magnetic 
field of $B\approx E_K /( h \mu_B)$ or approximately 100\,mG. We show in the next section that such a bias magnetic field will cause 
a significant Zeeman shift and cannot not be applied during clock operation. As mentioned in the Introduction, the temperature can be lowered to a few microkelvin which will reduce the threshold magnetic field to a few tens of milligauss which is sufficient for a target clock 
accuracy.

 Assuming that the lattice filling factor is less than unity and taking into account the influence of weak magnetic bias field 
 (10\,mG), we conclude that the spin relaxation time should be larger than 10\,ms. This sets the bound for the interrogation time of 
 the clock transition and, correspondingly, the Fourier limit of its spectral linewidth of  $<10$\,Hz. As a result, the spin 
 relaxation should not considerably impact the performance of the proposed optical clock.

\subsection{Interaction with an external magnetic field}
\label{sec:zeeman}

To selectively address  the $|m=0\rangle \rightarrow |m=0\rangle$ transition, an external static magnetic field $B$ has to be applied. The 
Hamiltonian describing hyperfine interaction for the $^{169}$Tm atom ($I=1/2$) in the presence of the external magnetic field $B$ 
is~\cite{Giglberger1967}:
\begin{equation}\label{eq:HFS}
H=h A \hat{\vec{I}}\cdot\hat{\vec{J}} - g_I \mu_N \hat{\vec{I}}\cdot \hat{\vec{B}}-g_J \mu_B \hat{\vec{J}}\cdot \hat{\vec{B}},
\end{equation}
where $A$ is the hyperfine constant, $g_I$ is the nuclear Land\'{e} g-factor,  $\mu_N$ is the nuclear magneton, and  $g_J$ is the electronic 
Land\'{e} g-factor. The well-known Breit-Rabi formula gives eigenvalues for the special case of $J=1/2$.
Making the formal substitution $I\leftrightarrow J$, $g_I \leftrightarrow g_J$, $\mu_N \leftrightarrow \mu_B$ in (\ref{eq:HFS})  one can 
use the Breit-Rabi expression for $I=1/2$~\cite{Giglberger1967}.
The frequency shift of the clock transition $|m=0\rangle \rightarrow |m=0\rangle$ is given by
\begin{multline}
\label{eq:quad_zeeman}
\varDelta f_{0 \rightarrow 0} = \frac{(g_{5/2}\mu_B - g_I\mu_N)^2 B^2}{4 h^2 \varDelta W_{5/2} } \\
-  \frac{(g_{7/2}\mu_B - g_I\mu_N)^2 B^2}{4 h^2\varDelta W_{7/2}} =\beta \, B^2
\end{multline}
and
\begin{equation}
\beta = -257(1)\text{\,Hz}/\text{G}^2,
\end{equation}
where $\varDelta W_{7/2} = 1496.550(1)$\,MHz, $\varDelta W_{5/2} =  2114.946(1)$\,MHz are the hyperfine frequency splittings of the $|J=7/2\rangle$ 
and $|J=5/2\rangle$ clock levels~\cite{Leeuwen1980},  and $g_{7/2} = 1.141189(3)$~\cite{Giglberger1967}, 
$g_{5/2}=0.855(1)$~\cite{Blaise1965} are their Land\'{e} g-factors, $g_I 
=  0.462(3)$~\cite{Giglberger1967}.

\section{ \texorpdfstring{T\MakeLowercase{m}}{Tm} clock uncertainty}
\label{sec:budget}
Here we will discuss the most significant sources of uncertainty for the proposed  Tm clock.
\subsection{Black body radiation}
\label{subsec:BBR}

The frequency shift of the clock transition due to the AC-Stark shift induced by BBR is given by
\begin{multline}
\varDelta f_{BBR}  =
 \int \limits_{\omega=0}^{\infty} \frac{ a_0^3\omega^3}{\pi^2 c^2} \frac{\left( \alpha_{gr}^s (\omega) - \alpha_{cl}^s(\omega) 
 \right)}{e^{\frac{\hbar \omega}{k_B T}-1}} d\omega \\
\approx \varDelta \alpha^s_0 \frac{a_0^3\pi^2 k_B^4}{15 c^3 \hbar^4}\, T^4 =
  1.17 \times 10^{-12} \, \varDelta \alpha^s_0 \text{[a.u.]}\,T^4\text{[K]}
\label{eq:mdd1}
\end{multline}
where $\varDelta \alpha^s_0$ is the differential scalar static polarizability of the clock levels in atomic units, $T$ is the temperature 
in kelvin.

Our  calculations (see sec.~\ref{sec:polarizability:theory}) gives  $\varDelta \alpha^s_0=2$\,a.u. which results in  $\varDelta f_{BBR} 
=20$\,mHz at $T=300$\,K. It corresponds to a fractional frequency shift of the clock transition of  $8\times 10^{-17}$  which is 
much less than for the Sr atom and is comparable to Al$^+$ clock transition~\cite{Mitroy2010}. Uncertainty of the ambient temperature of 
3\,K will introduce a frequency uncertainty  of $3\times10^{-18}$ (0.8\,mHz). 
Since there are no strong transitions from the clock levels in the infrared region  the dynamic BBR 
shift is negligibly small~\cite{middelmann2011}.

\subsection{Second order Zeeman shift}
\label{sec:quad_zeeman}
According  to (\ref{eq:quad_zeeman}), the frequency shift of the clock transition in the external magnetic field of $B=10$\,mG 
corresponds to  $-25.7(1)$\,mHz or $10^{-16}$ in fractional units. One can accurately measure the bias field $B$ by monitoring the  
Zeeman shift  of the  $|m=-4\rangle  \rightarrow |m=-3\rangle$ and $|m=4\rangle \rightarrow |m=3\rangle$ transitions~\cite{Rosenband2007}. The frequency splitting of 
these magnetic sensitive transitions in a magnetic field is equal to 
$\xi = 2(4 g_{J=7/2,F=4} - 3 g_{J=5/2,F=3})\mu_B/h = 6.00(2)$ \,MHz/G, where $g_{J=7/2,F=4}$ and $g_{J=5/2,3=4}$ are the Land\'{e} g-factors of the states $|J=7/2, F = 4 \rangle$ and $|J=5/2, F=3\rangle$, respectively.
Given that the linewidths of both transitions are smaller than $\delta f_{4 \rightarrow 3} = 100$\,Hz (the broadening due to the magnetic interaction (\ref{eq:mdd1}) is included), the bias magnetic field $B_0$ can 
be measured {\it{in situ}} with the uncertainty of $\varDelta B = \delta f_{4 \rightarrow 3}  / \xi \le 0.1$\,mG. Since the magnetic field 
can be  stabilized at the same level  over the interrogation sequence~\cite{ref:IonOpticalClocks:NIST:Science}, we take 0.1\,mG 
as an upper limit for the bias magnetic field instability and estimate the quadratic Zeeman shift's contribution  as $2\beta  B_0 \varDelta B_0 
= 0.5$\,mHz ($2\times 10^{-18}$ in fractional units) after correction.

\subsection{Dynamic light shifts}
\label{sec:lightshifts}

Fluctuations $\delta I$ of the laser intensity cause shifts and broadening of the clock transition originated  from a non-zero differential 
hyperpolarizability $\varDelta\gamma$:
\begin{equation}
\label{eq:hypererror}
\varDelta f_I = -\frac{\delta I}{I}\left(\frac{\varDelta \alpha}{4} I  + 2 \frac{\varDelta\gamma}{64} I^2 \right)= -\frac{\delta I}{I}  
\frac{\varDelta\gamma}{64} I^2\,,
\end{equation}
where we take into account that
\begin{equation}
 \frac{\varDelta \alpha}{4} I  + \frac{\varDelta\gamma}{64} I^2 =0
 \end{equation}
  at the magic wavelength.
Previously, we estimated that  $\varDelta\gamma/64 \times I^2 $ is less than $0.5$\,Hz for the given lattice parameters (see sec.\,\ref{sec:hyper}). 
Stabilizing the laser intensity at the level of $10^{-3}$ the uncertainty in the frequency of the clock transition can be reduced to 
0.5\,mHz or $2\times 10^{-18}$ in fractional units.

\subsection{Van der Waals and quadrupole interactions}
The electrostatic van der Waals interaction between two neutral atoms shifts the clock frequency by
$
 -(C_{6} a_B^6 E_H)/(h \,r^6),
$
where $C_6$ is the van der Waals coefficient in atomic units,  $a_B$ is the Bohr radius, and $E_H$ is the  Hartree energy.
Following~\cite{Kotochigova2011}, we estimated $C_6 \sim 6000$\,a.u. for $|J=7/2\rangle$ level.
For an atomic separation of $r=400$\,nm (atoms are placed in the 800-nm optical lattice, less than one atom per site) the 
van der Waals frequency shift  is less than  0.1\,mHz which corresponds to $4\times 10^{-19}$ in fractional units.

To estimate the contribution from the  quadrupole-quadrupole interaction, we calculated the quadruple moment for the ground state of Tm atom 
using the  COWAN package.  The result is $D\sim 0.5\,e a_B^2$ ($e$ is the elementary charge). The corresponding frequency shift is $D^2 
/ (4\pi \epsilon_0 r^5 h) < 0.1$\,mHz.

\subsection{Line pulling and geometrical effects}
In an external bias magnetic field of $B=10$\,mG, the $|J=7/2,F=4\rangle \rightarrow |J=5/2,F=3\rangle$ transition will be split into magnetic 
components. The line pulling effect~\cite{Marchi1984} can perturb the magnetic-insensitive clock $|m=0\rangle \rightarrow |m=0\rangle$ transition. 
Imperfect co-alignment of the magnetic field $B$ and the polarization of the interrogating laser beam (Fig.\,\ref{latticeGeometry}) leads to 
excitation of $|m=0\rangle \rightarrow |m=\pm1\rangle$ transitions  and also can cause the line pulling effect.

In both cases, the separation from the clock transition  to the nearest transition is not less than $10^3\gamma \approx $ 20\,kHz, 
where $\gamma = 20$\,Hz is the upper bound for the expected transition linewidth.  The corresponding incoherent line pulling  is 
negligible   ($<10^{-8}$\,Hz) and does not impact the clock performance.
For reading the clock transition, absorption spectroscopy is typically used and  we do not expect a contribution from the coherent line 
pulling ~\cite{Beyer2015}.

Another systematic effect related to the geometry can come from misalignment of the lattice light polarization and the  bias magnetic 
field (see Fig.\,\ref{latticeGeometry}). The shift results from the differential tensor polarizability of the clock levels  $\varDelta 
\alpha^t$ and scales as the square of the misalignment angle~\cite{Nicholson2015}. It was shown that the corresponding relative 
frequency shift can be reduced to less than $2\times 10^{-18}$ by proper alignment~\cite{Nicholson2015}. 

\subsection{Uncertainty budget}
\label{sec:Uncertainty}

\renewcommand*{\arraystretch}{1.4}
\begin{table}[t]
\caption{ Uncertainty budget for a Tm optical clock operating at the 1.14 $\mu$m magnetic dipole transition. Atoms are trapped in the 
optical lattice at the magic wavelength close to 807\,nm with the light intensity of 50\,kW/cm$^2$.}
\label{budget}
\begin{tabular}{>{\centering\arraybackslash}b{3.1cm} >{\raggedleft\arraybackslash}b{1.5cm} >{\raggedleft\arraybackslash}b{1.8cm}  >{\raggedleft\arraybackslash}b{1.8cm} }
\hline\hline
Contribution	&	Frequency shift, mHz 	&	Uncertainty after correc- tion, mHz	&  Uncertainty in fractional units, $10^{-18}$\\
\hline
BBR ($T=300\pm3$\,K)																						&	20				&	0.8			& 3 		\\
Zeeman shift ($B=10.0\pm0.1$\,mG)	  			  									   &	$-26$		&	0.5			& 2		\\
Light shift due to hyperpolarizability	($\delta I/I=10^{-3}$)		&		0				&	0.5			& 2		\\
Light shift due to tensor polarizability														&	0.5			 	&	0.5 		& 2	 	\\
van der Waals	and quadrupole  interaction										&	0.1				& 	0.1			& 0.4	\\
\hline
Total																																	&	$-6	$		&		1.2		& $<5$\\
\hline\hline
\end{tabular}
\end{table}

The list of dominant frequency shifts and corresponding uncertainties is presented in Table\,\ref{budget}. The major line 
shifts are the BBR shift  and the second-order Zeeman shift. All of these can be well characterized and corrected to a high degree 
using moderate assumptions and established experimental techniques. Light shift can also be controlled at low $10^{-18}$ level by 
intensity stabilization of the light field. As a result, the systematic frequency uncertainty of the proposed Tm optical clock at 
$1.14$\,$\mu$m can be reduced to  $5 \times 10^{-18}$ in fractional units.

\section{Experiment}
\label{sec:experiment}
The experimental section describes our measurement of the $|J=5/2\rangle$ clock level lifetime in a dilute cloud of cold Tm atoms.   
Formerly, the decay from this level was studied in Tm atoms implanted in solid and liquid $^4$He \cite{ishikawa1997}. Strong shielding of 
inner shells and the high symmetry of the perturbing field of the He matrix give the impressive result of 75(3)\,ms for the 
lifetime of the $|J=5/2\rangle$ clock level. Note that the level was populated by a cascade decay from highly excited levels.

In contrast to \cite{ishikawa1997}, we directly excite the $|J=7/2,F=4\rangle \rightarrow |J=5/2,F=3\rangle$ transition  by spectrally narrow laser radiation at 1.14\,$\mu$m in an ensemble of Tm atoms trapped in a 1D optical lattice and measure the lifetime of the $|J=5/2\rangle$ clock level monitoring its decay to the ground $|J=7/2,F=4\rangle$ state.
Besides that, we  evaluated dynamic polarizability of Tm atoms at  532\,nm by exciting parametric resonances in an optical lattice.

\subsection{Lifetime of the $|J=5/2\rangle$ clock level}
\label{sec:lifetime}
 The lifetime of  $|J=5/2\rangle$ clock level is measured by excitation of the magnetic dipole transition at 1.14\,$\mu$m in a 1D optical 
 lattice. About $10^6$ thulium atoms are laser cooled down to 20\,$\mu$K in a narrow-line MOT operating at 
 530.7\,nm~\cite{Sukachev2014} and then are recaptured by the 1D optical lattice. The  lattice  is formed by a retro-reflected focused 
 532\,nm cw laser beam (waist radius is 50\,$\mu$m, laser power is 3\,W) and superimposed with the atomic cloud. The trap 
 depth is calculated to be 400\,$\mu$K for the ground $|J=7/2,F=4\rangle$ state which provides a recapture efficiency of 
 40\%~\cite{Sukachev2014}.

\begin{figure}[t]
\includegraphics[width=1\linewidth]{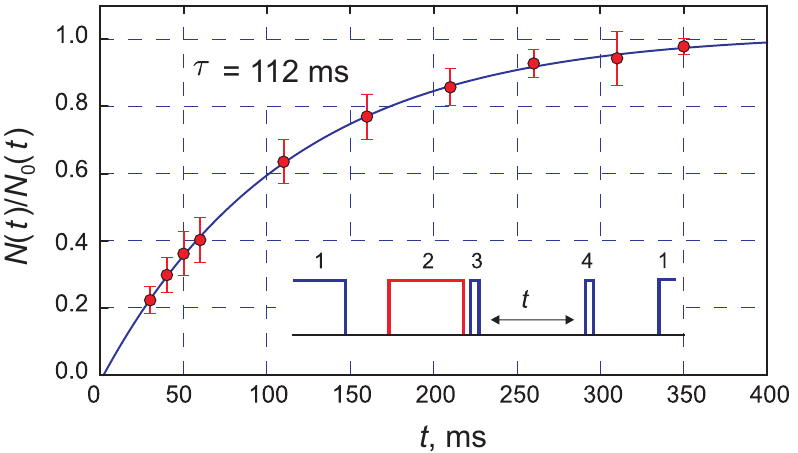}
\caption{\label{fig_lifetime} Measurement of the lifetime of the $|J=5/2\rangle$ level. Red dots are the normalized number of atoms decayed to the ground $|J=7/2,F=4\rangle$ level. 
The solid curve is a fit by (\ref{eq:exp}).
The inset shows the pulse sequence described in the text.
}
\end{figure}

After recapture (pulse 1 in Fig.\,\ref{fig_lifetime}), we switch  the  MOT off and  wait for 20\,ms to let uncaptured atoms escape. 
Then, a resonant  1.14\,$\mu$m laser pulse of 30\,ms (pulse 2) is applied to excite atoms to the $|J=5/2,F=3\rangle$ 
level~\cite{Golovizin2015}. The laser is actively stabilized to a high-finesse ULE cavity~\cite{Alnis2008} which narrows the laser 
spectral linewidth down to $\sim 10$\,Hz. After the  interrogation pulse, a resonant 410.6\,nm laser pulse of 1\,ms (pulse 3) is 
applied to remove atoms from the $|J=7/2\rangle$ ground state (Fig.~\ref{fig_levels}).
Atoms exited to the $|J=5/2,F=3\rangle$  decay back to the $|J=7/2,F=4\rangle$ ground state, which population is monitored by a fluorescence signal induced by a delayed 410.6\,nm  probe pulse (pulse 4).

The increase of the population of the $|J=7/2,F=4\rangle$ ground state is  described by the  exponential function
\begin{equation}
 N(t) = N_{0}(1-\exp(-{t}/{\tau})),
\label{eq:exp}
\end{equation}
where $\tau$ is the lifetime of the the excited  $|J=5/2,F=3\rangle$ state and $N_{0}$ is the initial number of atoms in this state.  By fitting the experimental data presented in  
Fig.~\ref{fig_lifetime}, we measure $\tau=112(4)$~ms.  It is the lower bound for the $|J=5/2\rangle$ level natural lifetime since the measured lifetime  can be reduced by  additional weak losses from the $|J=5/2\rangle$ level in the optical lattice. These losses may be related to optical or magnetic Feshbach resonances~\cite{Chin2010}.

Thus, the natural linewidth of the clock transition $|J=7/2\rangle \rightarrow |J=5/2\rangle$  is expected to be not broader than  1.4\,Hz which is consistent with 
the previous measurement in $^4$He matrix \cite{ishikawa1997} and the theoretical prediction of 1.14\,Hz~\cite{Kolachevsky2007}. The 
natural linewidth of the transition does not limit the performance of the proposed optical clock (see sec.~\ref{sec:budget}), because for  most routinely operating optical clocks the Fourier-limited spectral linewidth of the clock transition is on the order of 10\,Hz.

In the current experimental arrangement we observed the spectral linewidth of the $|J=7/2,F=4\rangle \rightarrow |J=5/2,F=3\rangle$ transition of 1\,MHz at the low power limit~\cite{Vishnyakova2016}.  It is due to Zeeman splitting in the un-compensated laboratory field ($\sim 0.5$\,MHz) and  inhomogeneous power broadening caused by different dynamic polarizabilities of the $|J=7/2,F=4\rangle$ and $|J=5/2,F=3\rangle$ levels at 532\,nm ($\sim 0.4$\,MHz).

\subsection{Parametric resonances}
\label{sec:polarizability532nm}
The dynamic  polarizability  of the $|J=7/2\rangle$ ground state at the lattice wavelength can be evaluated by the excitation of parametric 
resonances in the lattice and by monitoring the corresponding losses~\cite{Friebel1998}. This method is very sensitive to the laser 
beam parameters (waist size, astigmatism) and does not allow accurate comparison to our calculations. Still, it gives the proper 
order of magnitude for the polarizability and provides unambiguous proof that Tm atoms are localized in the 1D optical lattice.

At low temperatures, atomic motion in the optical lattice becomes  quantized and the corresponding  axial $f_a$ and  radial $f_r$ oscillation 
frequencies at the center of the lattice are given by
\begin{equation}
\begin{split}
\label{eq:parametric0}
f_a =& \frac{4}{w_0 \lambda}\sqrt{\frac{2  a_0^3 \alpha^s P}{c m_0 }}\,,\\
f_r =&\frac{4}{\pi w_0^2} \sqrt{\frac{a_0^3 \alpha^s P}{c  m_0}}\,,
\end{split}
\end{equation}
where $P=4$\,W is the optical power of the laser beam forming the 1D lattice, $m_0$ is the Tm atomic  mass, $w_0$ is the beam waist radius (at 
$1/e^2$ intensity level), $\lambda=532$\,nm is the lattice wavelength, and $\alpha^s$ is the scalar polarizability at $\lambda=532$\,nm of the $|J=7/2,F=4\rangle$ level in atomic units. 
According to~\cite{Landau:Mechanics}, harmonic modulation of the trap depth at  frequencies $2f / n$ (here $f$ is one of the 
eigenfrequencies (\ref{eq:parametric0}) and $n$ is an integer) will cause parametric excitation of the resonances and corresponding 
trap losses.

\begin{figure}
\includegraphics[width=1\linewidth]{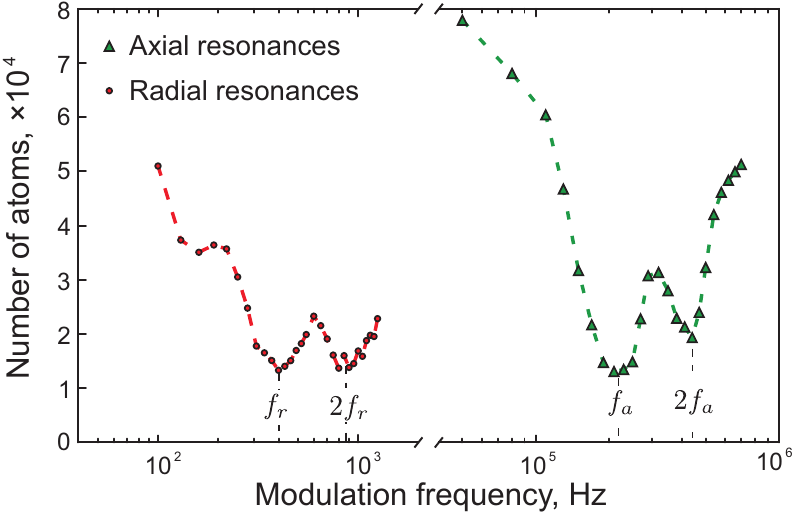}
\caption{Excitation of the parametric resonances in the 532\,nm 1D optical lattice. Green triangles are axial resonances, red circles are radial ones; dashed curves are guides to the eyes.}
\label{fig:parametric}
\end{figure}

To excite parametric resonances in the 532\,nm optical lattice, we harmonically modulated the  laser power and, correspondingly, the 
trap depth by an acousto-optical modulator (AOM) at the level of 10\%. The number of atoms remaining in the optical lattice after 
100\,ms of parametric excitation was monitored by  resonance fluorescence at 410.6\,nm. The corresponding spectrum  is shown in  
Fig.~\ref{fig:parametric}. The low frequency parametric resonances at  $400(40)$\,Hz and $900(150)$\,Hz correspond to the radial 
oscillations at $f_r$ and $2f_r$ frequencies. The high frequency resonances at $230(40)$\,kHz and $420(50)$\,kHz are related to axial 
oscillations at $f_a$ and $2f_a$  in the tight potential wells of the lattice. Higher order parametric resonances are much weaker and 
broader~\cite{Landau:Mechanics} and were not observed.

The scalar polarizability can be deduced from  (\ref{eq:parametric0}) by excluding  $w_0$:
\begin{equation}
\alpha^s = \frac{f_a^4 \lambda^4 c  m_0}{64 f_r^2 a_0^3 \pi^2 P}\,.
\end{equation}
From the measured frequencies we estimate this value to be $360^{+300}_{-200}$ a.u. which agrees with the calculated 
polarizability of 600\,a.u.  within  error bars. The main sources of uncertainty are astigmatism in the lattice beams, axial and radial 
misalignment of the waist positions of the lattice beams, and error in our determination of the parametric resonance frequency.

In conclusion, the experimental results for the scalar dynamic polarizability  $\alpha^s$ at 532\, nm  in the optical lattice and in 
the dipole trap are consistent  with the calculated  value of 600\,a.u. Although the experiment does not allow to test the accuracy of 
our calculations, it unambiguously proves trapping Tm atoms in the optical lattice at 532\,nm.


\section{Summary}
We considered the possibility to use the inner-shell transition $[\text{Xe}]4f^{13}6s^2 (J=7/2,\,F=4,\,m=0)  \rightarrow [\text{Xe}]4f^{13}6s^2 
(J=5/2,\,F=3,\,m=0)$ in the Tm atom at $\lambda=1.14\,\mu$m as a candidate for an optical lattice clock. The transition wavelengths and 
probabilities for two clock levels $|J=7/2\rangle$ and $|J=5/2\rangle$ in the spectral range 250\,nm --- 1200\,nm are calculated using the COWAN 
package which allows deducing of the  differential dynamic polarizability and suggests that the magic wavelength is at around 807\,nm with 
an attractive optical potential. Our calculations show a reasonable correspondence with existing experimental data and significantly extend
it to the UV and IR spectral ranges.

The suggested clock transition demonstrates a low sensitivity to the BBR shift which provides a clock frequency instability at the low $10^{-18}$ 
level competing with the best known optical clocks. We also evaluated other feasible contributions to clock performance (magnetic 
interactions, light shifts, van der Waals, and quadrupole shifts) which, after reasonable assumptions, can be lowered to the $10^{-18}$ 
level. Together with the relative simplicity of laser cooling and trapping Tm atoms, our results demonstrate that Tm is a promising candidate for optical clock 
applications. One of the disadvantages is the relatively low carrier frequency of only $2.6\times10^{14}$\,Hz which requires longer 
integration time to reach the same instability as  Sr and Yb lattice clocks.

Our experiments with direct excitation of the clock transition by spectrally narrow laser radiation at $\lambda=1.14\,\mu$m set a 
lower limit for the upper clock level lifetime of 112\,ms which corresponds to the natural linewidth of $<1.4$\,Hz. Experiments were 
done in a 1D  optical lattice  at 532\,nm. Modulating the trap depth and analyzing the  
corresponding parametric resonances frequencies, we deduced the scalar polarizability of the Tm ground state at 532\,nm which shows reasonable agreement with our calculations.

To experimentally study the magic wavelength and analyze systematic shifts, we plan to change the trapping wavelength to 806\,nm --- 
807\,nm using a tunable Ti:sapphire laser. This will also simplify our study of Feshbach resonances and may open a way to study and 
control dipole-dipole interactions using narrow band excitation of the clock transition at 1.14\,nm.
\begin{acknowledgments}
 The work is supported by RFBR grants \#15-02-05324 and \#16-29-11723. We are grateful to S.\,Kanorski and V.\,Belyaev for invaluable 
 technical support.
\end{acknowledgments}

\appendix

\section{Transition probabilities}
\label{sec:calculated_probabilities}

\begin{figure}[t]
\includegraphics[width=1\linewidth]{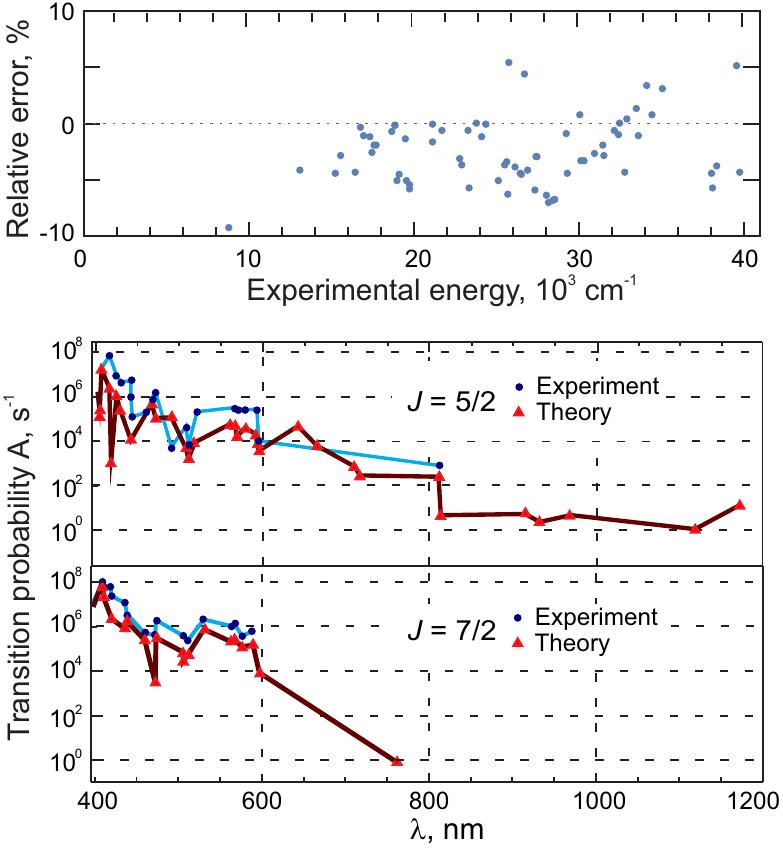}
\caption{Comparison between the calculated and measured data. Only levels which were identified with the experimentally measured ones are shown.
Upper panel: Relative error of calculated energy levels.  
Lower panel: Calculated (red triangles) and measured (blue circles) transition probabilities from the upper $|J=5/2\rangle$ and the lower $|J=7/2\rangle$ clock levels. Solid curves are guides for the eyes only.}
\label{fig:COWAN}
\end{figure}

\begin{table}[b]
\caption{Calculated (experimentally unmeasured) transition probabilities from the $|J=7/2\rangle$ and $|J=5/2\rangle$ clock levels. $E$ and $J$ are experimentally measured energies and electronic angular momenta of the upper levels of the transitions, respectively.}
\label{table:cowan}
\begin{tabular}{l c >{\raggedleft\arraybackslash}p{1.5cm} >{\raggedleft\arraybackslash}p{1.8cm} }
\hline\hline
$E$, $10^3$\,cm$^{-1}$	&	J	&	$\lambda$, nm	 & $A$ c$^{-1}$\\
\hline\hline
\multicolumn{4}{c}{$|J=7/2\rangle$ level}\\
\hline
 $13\,119.6$	&	$ \sfrac{9}{2} $	&	$ 762.22 $	&	$ 3.2 $\\
 $16\,742.2 $	&	$\sfrac{7}{2} $	&	$ 597.29 $	&	$ 1.9\times10^5 $\\
 $19\,748.5 $	&	$ \sfrac{9}{2} $	&	$ 506.37 $	&	$ 1.2\times10^6  $\\
\hline
\multicolumn{4}{c}{$|J=5/2\rangle$ level}\\
\hline
	$16\,742.2$	    &	$ \sfrac{7}{2} $	&	$ 1254.55  $	&	$ 14.1 $\\
	$16\,957.0 $	&	$ \sfrac{7}{2} $ 	&	$ 1221.63 $	&	$ 1.2 $\\
	$17\,343.4 $	&	$ \sfrac{7}{2} $	&	$ 1166.57 $	&	$ 171.2 $\\
	$17\,752.6 $	&	$ \sfrac{5}{2} $	&	$ 1113.41 $	&	$ 27.5 $\\
	$19\,132.2 $	&	$ \sfrac{3}{2} $	&	$ 965.16 $	&	$ 150 $\\
	$19\,548.8 $	&	$ \sfrac{5}{2} $	&	$ 927.85 $	&	$ 100$\\
	$19\,753.8 $	&	$ \sfrac{7}{2} $	&	$ 910.53 $	&	$ 209 $\\
	$21\,120.8 $	&	$ \sfrac{7}{2} $	&	$ 809.74 $	&	$ 640 $\\
	$22\,791.2 $	&	$ \sfrac{7}{2} $	&	$ 713.27 $	&	$ 27\,400 $\\
	$22\,929.7 $	&	$ \sfrac{5}{2} $	&	$ 706.29 $	&	$ 58\,900 $\\
	$23\,873.2 $	&	$ \sfrac{7}{2} $	&	$  662.17$	&	$ 1\times10^6 $\\
	$23\,882.4 $	&	$ \sfrac{3}{2} $	&	$ 661.76 $	&	$ 2.13\times10^6 $\\
	$24\,418.4 $	&	$ \sfrac{5}{2} $	&	$ 639.11 $	&	$ 1.79\times10^7$\\
\hline\hline
\end{tabular}
\end{table}

Calculation of  dynamic polarizabilities eq.\,(\ref{eq:alphaFM}-\ref{eq:alphaJFt}) requires knowledge of electrical dipole transition
rates from the lower $|J=7/2\rangle$  and the upper $|J=5/2\rangle$ clock levels in Tm. A number of transitions rates were experimentally measured and 
the results are
summarized in~\cite{TmTransitions}.  We completed this list by calculation of energy levels in the range up to 40000\,cm$^{-1}$ and 
corresponding transition-dipole matrix elements using the COWAN package \cite{Cowan1981} taking into account the low lying odd 
($4f^{13}6s^2$, $4f^{13} 5d^1 6s^1$, $4f^{12} 6s^2 6p^1$, $4f^{13}
6p^2$, $4f^{13} 5d^2$) and even ($4f^{12} 5d^1 6s^2$, $4f^{13} 6s^1 6p^1$) configurations.

As follows from the upper panel of Fig.\,\ref{fig:COWAN}, the accuracy of the calculated level energies is better than 20\%. Together with 
the known leading configuration percentage it is sufficient to identify the low laying levels  with the energies $E <20000$\,cm$^{-1}$ ($\lambda > 500$\,nm) and 
compare them with the experimentally measured ones~\cite{TmEnergyLevels}. Table\,\ref{table:cowan} shows 
calculated probabilities of the transitions  with experimentally unknown probabilities from the clock levels  $[\text{Xe}]4f^{13}6s^2 (J=7/2)$ 
and $[\text{Xe}]4f^{13}6s^2 (J=5/2)$.  The lower panel of Fig.\,\ref{fig:COWAN} shows comparison between the experimentally measured and 
calculated transition probabilities in the range from  500\,nm to 1200\,nm where level identification can be done unambiguously. Taking into 
account difficulties with simulation of the hollow-shell atomic potentials~\cite{Lepers2014}, the discrepancy between the calculated and 
the experimental data seems to be reasonable.

As mentioned in the main text, the most self-consistent approach for deriving differential polarizability of the clock levels is to 
use the calculated data for the transition probabilities, otherwise we meet difficulties with level identification in the wavelength range 
$\lambda<500$\,nm and with matching the calculated and  the experimental data. In turn, transitions to the  highly excited Rydberg states 
($\lambda<250$\,nm) become extremely dense and COWAN package cannot be used. To calculate their contribution one typically uses 
semi-analytical approach~\cite{Chernov2005}. In our case, strong similarity of spectra 
starting from two inner-shell fine structure sublevels -- clock levels ($|J=7/2\rangle$ and $|J=5/2\rangle$) -- results in very small differential polarizability of 0.1\,a.u. if one takes into account all transitions in the spectral range 250\,nm --- 1200\,nm. Transitions to the Rydberg  states may 
slightly influence the absolute values of the polarizability of the clock states (at the level of a few a.u.), but we do not expect 
significant contribution to the differential polarizability $\varDelta\alpha$.

\bibliography{polarizability}

\begin{thebibliography}{60}%
\makeatletter
\providecommand \@ifxundefined [1]{%
 \@ifx{#1\undefined}
}%
\providecommand \@ifnum [1]{%
 \ifnum #1\expandafter \@firstoftwo
 \else \expandafter \@secondoftwo
 \fi
}%
\providecommand \@ifx [1]{%
 \ifx #1\expandafter \@firstoftwo
 \else \expandafter \@secondoftwo
 \fi
}%
\providecommand \natexlab [1]{#1}%
\providecommand \enquote  [1]{``#1''}%
\providecommand \bibnamefont  [1]{#1}%
\providecommand \bibfnamefont [1]{#1}%
\providecommand \citenamefont [1]{#1}%
\providecommand \href@noop [0]{\@secondoftwo}%
\providecommand \href [0]{\begingroup \@sanitize@url \@href}%
\providecommand \@href[1]{\@@startlink{#1}\@@href}%
\providecommand \@@href[1]{\endgroup#1\@@endlink}%
\providecommand \@sanitize@url [0]{\catcode `\\12\catcode `\$12\catcode
  `\&12\catcode `\#12\catcode `\^12\catcode `\_12\catcode `\%12\relax}%
\providecommand \@@startlink[1]{}%
\providecommand \@@endlink[0]{}%
\providecommand \url  [0]{\begingroup\@sanitize@url \@url }%
\providecommand \@url [1]{\endgroup\@href {#1}{\urlprefix }}%
\providecommand \urlprefix  [0]{URL }%
\providecommand \Eprint [0]{\href }%
\providecommand \doibase [0]{http://dx.doi.org/}%
\providecommand \selectlanguage [0]{\@gobble}%
\providecommand \bibinfo  [0]{\@secondoftwo}%
\providecommand \bibfield  [0]{\@secondoftwo}%
\providecommand \translation [1]{[#1]}%
\providecommand \BibitemOpen [0]{}%
\providecommand \bibitemStop [0]{}%
\providecommand \bibitemNoStop [0]{.\EOS\space}%
\providecommand \EOS [0]{\spacefactor3000\relax}%
\providecommand \BibitemShut  [1]{\csname bibitem#1\endcsname}%
\let\auto@bib@innerbib\@empty
\bibitem [{\citenamefont {{E V Zharikov, V I Zhekov, L A Kulevskii, T M
  Murina}}\ \emph {et~al.}(1975)\citenamefont {{E V Zharikov, V I Zhekov, L A
  Kulevskii, T M Murina}}, \citenamefont {Osiko}, \citenamefont {Prokhorov},
  \citenamefont {Savel'ev}, \citenamefont {{V V Smirnov}}, \citenamefont {{B P
  Starikov}},\ and\ \citenamefont {{M I Timoshechkin}}}]{Zharkov1975}%
  \BibitemOpen
  \bibfield  {author} {\bibinfo {author} {\bibfnamefont {V.~V.}\ \bibnamefont
  {{E V Zharikov, V I Zhekov, L A Kulevskii, T M Murina}}}, \bibinfo {author}
  {\bibfnamefont {A.~M.}\ \bibnamefont {Osiko}}, \bibinfo {author}
  {\bibfnamefont {A.~D.}\ \bibnamefont {Prokhorov}}, \bibinfo {author}
  {\bibnamefont {Savel'ev}}, \bibinfo {author} {\bibnamefont {{V V Smirnov}}},
  \bibinfo {author} {\bibnamefont {{B P Starikov}}}, \ and\ \bibinfo {author}
  {\bibnamefont {{M I Timoshechkin}}},\ }\href {\doibase
  10.1070/QE1975v004n08ABEH011147} {\bibfield  {journal} {\bibinfo  {journal}
  {Sov J Quantum Electron}\ }\textbf {\bibinfo {volume} {4}},\ \bibinfo {pages}
  {1039} (\bibinfo {year} {1975})}\BibitemShut {NoStop}%
\bibitem [{\citenamefont {Barnes}\ \emph {et~al.}(1993)\citenamefont {Barnes},
  \citenamefont {Filer}, \citenamefont {Naranjo}, \citenamefont {Rodriguez},\
  and\ \citenamefont {Kokta}}]{Barnes1993}%
  \BibitemOpen
  \bibfield  {author} {\bibinfo {author} {\bibfnamefont {N.~P.}\ \bibnamefont
  {Barnes}}, \bibinfo {author} {\bibfnamefont {E.~D.}\ \bibnamefont {Filer}},
  \bibinfo {author} {\bibfnamefont {F.~L.}\ \bibnamefont {Naranjo}}, \bibinfo
  {author} {\bibfnamefont {W.~J.}\ \bibnamefont {Rodriguez}}, \ and\ \bibinfo
  {author} {\bibfnamefont {M.~R.}\ \bibnamefont {Kokta}},\ }\href {\doibase
  10.1364/OL.18.000708} {\bibfield  {journal} {\bibinfo  {journal} {Optics
  Letters}\ }\textbf {\bibinfo {volume} {18}},\ \bibinfo {pages} {708}
  (\bibinfo {year} {1993})}\BibitemShut {NoStop}%
\bibitem [{\citenamefont {B{\"{o}}ttger}\ \emph {et~al.}(2001)\citenamefont
  {B{\"{o}}ttger}, \citenamefont {Pryde}, \citenamefont {Strickland},
  \citenamefont {Sellin},\ and\ \citenamefont {Cone}}]{Boettger2001}%
  \BibitemOpen
  \bibfield  {author} {\bibinfo {author} {\bibfnamefont {T.}~\bibnamefont
  {B{\"{o}}ttger}}, \bibinfo {author} {\bibfnamefont {G.~J.}\ \bibnamefont
  {Pryde}}, \bibinfo {author} {\bibfnamefont {N.~M.}\ \bibnamefont
  {Strickland}}, \bibinfo {author} {\bibfnamefont {P.~B.}\ \bibnamefont
  {Sellin}}, \ and\ \bibinfo {author} {\bibfnamefont {R.~L.}\ \bibnamefont
  {Cone}},\ }\href {\doibase 10.1364/OPN.12.12.000023} {\bibfield  {journal}
  {\bibinfo  {journal} {Optics and Photonics News}\ }\textbf {\bibinfo {volume}
  {12}},\ \bibinfo {pages} {23} (\bibinfo {year} {2001})}\BibitemShut {NoStop}%
\bibitem [{\citenamefont {Aleksandrov}\ \emph {et~al.}(1983)\citenamefont
  {Aleksandrov}, \citenamefont {Kotylev}, \citenamefont {Kulyasov},\ and\
  \citenamefont {Vasilevskii}}]{ref:AlexandrovEnglish1983}%
  \BibitemOpen
  \bibfield  {author} {\bibinfo {author} {\bibfnamefont {E.}~\bibnamefont
  {Aleksandrov}}, \bibinfo {author} {\bibfnamefont {V.}~\bibnamefont
  {Kotylev}}, \bibinfo {author} {\bibfnamefont {V.}~\bibnamefont {Kulyasov}}, \
  and\ \bibinfo {author} {\bibfnamefont {K.}~\bibnamefont {Vasilevskii}},\
  }\href@noop {} {\bibfield  {journal} {\bibinfo  {journal} {Opt. Spektrosk.}\
  }\textbf {\bibinfo {volume} {54}},\ \bibinfo {pages} {3} (\bibinfo {year}
  {1983})}\BibitemShut {NoStop}%
\bibitem [{\citenamefont {Wilpers}\ \emph {et~al.}(2002)\citenamefont
  {Wilpers}, \citenamefont {Binnewies}, \citenamefont {Degenhardt},
  \citenamefont {Sterr}, \citenamefont {Helmcke},\ and\ \citenamefont
  {Riehle}}]{Wilpers2002}%
  \BibitemOpen
  \bibfield  {author} {\bibinfo {author} {\bibfnamefont {G.}~\bibnamefont
  {Wilpers}}, \bibinfo {author} {\bibfnamefont {T.}~\bibnamefont {Binnewies}},
  \bibinfo {author} {\bibfnamefont {C.}~\bibnamefont {Degenhardt}}, \bibinfo
  {author} {\bibfnamefont {U.}~\bibnamefont {Sterr}}, \bibinfo {author}
  {\bibfnamefont {J.}~\bibnamefont {Helmcke}}, \ and\ \bibinfo {author}
  {\bibfnamefont {F.}~\bibnamefont {Riehle}},\ }\href {\doibase
  10.1103/PhysRevLett.89.230801} {\bibfield  {journal} {\bibinfo  {journal}
  {Physical Review Letters}\ }\textbf {\bibinfo {volume} {89}},\ \bibinfo
  {pages} {230801} (\bibinfo {year} {2002})}\BibitemShut {NoStop}%
\bibitem [{\citenamefont {Ido}\ \emph {et~al.}(2005)\citenamefont {Ido},
  \citenamefont {Loftus}, \citenamefont {Boyd}, \citenamefont {Ludlow},
  \citenamefont {Holman},\ and\ \citenamefont {Ye}}]{Ido2005}%
  \BibitemOpen
  \bibfield  {author} {\bibinfo {author} {\bibfnamefont {T.}~\bibnamefont
  {Ido}}, \bibinfo {author} {\bibfnamefont {T.~H.}\ \bibnamefont {Loftus}},
  \bibinfo {author} {\bibfnamefont {M.~M.}\ \bibnamefont {Boyd}}, \bibinfo
  {author} {\bibfnamefont {A.~D.}\ \bibnamefont {Ludlow}}, \bibinfo {author}
  {\bibfnamefont {K.~W.}\ \bibnamefont {Holman}}, \ and\ \bibinfo {author}
  {\bibfnamefont {J.}~\bibnamefont {Ye}},\ }\href {\doibase
  10.1103/PhysRevLett.94.153001} {\bibfield  {journal} {\bibinfo  {journal}
  {Physical Review Letters}\ }\textbf {\bibinfo {volume} {94}},\ \bibinfo
  {pages} {153001} (\bibinfo {year} {2005})}\BibitemShut {NoStop}%
\bibitem [{\citenamefont {Hancox}\ \emph {et~al.}(2004)\citenamefont {Hancox},
  \citenamefont {Doret}, \citenamefont {Hummon}, \citenamefont {Luo},\ and\
  \citenamefont {Doyle}}]{Hancox2004}%
  \BibitemOpen
  \bibfield  {author} {\bibinfo {author} {\bibfnamefont {C.~I.}\ \bibnamefont
  {Hancox}}, \bibinfo {author} {\bibfnamefont {S.~C.}\ \bibnamefont {Doret}},
  \bibinfo {author} {\bibfnamefont {M.~T.}\ \bibnamefont {Hummon}}, \bibinfo
  {author} {\bibfnamefont {L.}~\bibnamefont {Luo}}, \ and\ \bibinfo {author}
  {\bibfnamefont {J.~M.}\ \bibnamefont {Doyle}},\ }\href {\doibase
  10.1038/nature02938} {\bibfield  {journal} {\bibinfo  {journal} {Nature}\
  }\textbf {\bibinfo {volume} {431}},\ \bibinfo {pages} {281} (\bibinfo {year}
  {2004})}\BibitemShut {NoStop}%
\bibitem [{\citenamefont {Connolly}\ \emph {et~al.}(2010)\citenamefont
  {Connolly}, \citenamefont {Au}, \citenamefont {Doret}, \citenamefont
  {Ketterle},\ and\ \citenamefont {Doyle}}]{Connolly2010}%
  \BibitemOpen
  \bibfield  {author} {\bibinfo {author} {\bibfnamefont {C.~B.}\ \bibnamefont
  {Connolly}}, \bibinfo {author} {\bibfnamefont {Y.~S.}\ \bibnamefont {Au}},
  \bibinfo {author} {\bibfnamefont {S.~C.}\ \bibnamefont {Doret}}, \bibinfo
  {author} {\bibfnamefont {W.}~\bibnamefont {Ketterle}}, \ and\ \bibinfo
  {author} {\bibfnamefont {J.~M.}\ \bibnamefont {Doyle}},\ }\href {\doibase
  10.1103/PhysRevA.81.010702} {\bibfield  {journal} {\bibinfo  {journal}
  {Physical Review A}\ }\textbf {\bibinfo {volume} {81}},\ \bibinfo {pages}
  {010702} (\bibinfo {year} {2010})}\BibitemShut {NoStop}%
\bibitem [{\citenamefont {Katori}\ \emph {et~al.}(2003)\citenamefont {Katori},
  \citenamefont {Takamoto}, \citenamefont {Pal'chikov},\ and\ \citenamefont
  {Ovsiannikov}}]{Katori2003}%
  \BibitemOpen
  \bibfield  {author} {\bibinfo {author} {\bibfnamefont {H.}~\bibnamefont
  {Katori}}, \bibinfo {author} {\bibfnamefont {M.}~\bibnamefont {Takamoto}},
  \bibinfo {author} {\bibfnamefont {V.~G.}\ \bibnamefont {Pal'chikov}}, \ and\
  \bibinfo {author} {\bibfnamefont {V.~D.}\ \bibnamefont {Ovsiannikov}},\
  }\href {\doibase 10.1103/PhysRevLett.91.173005} {\bibfield  {journal}
  {\bibinfo  {journal} {Physical Review Letters}\ }\textbf {\bibinfo {volume}
  {91}},\ \bibinfo {pages} {173005} (\bibinfo {year} {2003})}\BibitemShut
  {NoStop}%
\bibitem [{\citenamefont {Takamoto}\ and\ \citenamefont
  {Katori}(2003)}]{Takamoto2003}%
  \BibitemOpen
  \bibfield  {author} {\bibinfo {author} {\bibfnamefont {M.}~\bibnamefont
  {Takamoto}}\ and\ \bibinfo {author} {\bibfnamefont {H.}~\bibnamefont
  {Katori}},\ }\href {\doibase 10.1103/PhysRevLett.91.223001} {\bibfield
  {journal} {\bibinfo  {journal} {Physical review letters}\ }\textbf {\bibinfo
  {volume} {91}},\ \bibinfo {pages} {223001} (\bibinfo {year}
  {2003})}\BibitemShut {NoStop}%
\bibitem [{\citenamefont {Ludlow}\ \emph {et~al.}(2015)\citenamefont {Ludlow},
  \citenamefont {Boyd}, \citenamefont {Ye}, \citenamefont {Peik},\ and\
  \citenamefont {Schmidt}}]{Ludlow2015}%
  \BibitemOpen
  \bibfield  {author} {\bibinfo {author} {\bibfnamefont {A.~D.}\ \bibnamefont
  {Ludlow}}, \bibinfo {author} {\bibfnamefont {M.~M.}\ \bibnamefont {Boyd}},
  \bibinfo {author} {\bibfnamefont {J.}~\bibnamefont {Ye}}, \bibinfo {author}
  {\bibfnamefont {E.}~\bibnamefont {Peik}}, \ and\ \bibinfo {author}
  {\bibfnamefont {P.~O.}\ \bibnamefont {Schmidt}},\ }\href {\doibase
  10.1103/RevModPhys.87.637} {\bibfield  {journal} {\bibinfo  {journal}
  {Reviews of Modern Physics}\ }\textbf {\bibinfo {volume} {87}},\ \bibinfo
  {pages} {637} (\bibinfo {year} {2015})}\BibitemShut {NoStop}%
\bibitem [{\citenamefont {Bloom}\ \emph {et~al.}(2014)\citenamefont {Bloom},
  \citenamefont {Nicholson}, \citenamefont {Williams}, \citenamefont
  {Campbell}, \citenamefont {Bishof}, \citenamefont {Zhang}, \citenamefont
  {Zhang}, \citenamefont {Bromley},\ and\ \citenamefont {Ye}}]{Bloom2014}%
  \BibitemOpen
  \bibfield  {author} {\bibinfo {author} {\bibfnamefont {B.~J.}\ \bibnamefont
  {Bloom}}, \bibinfo {author} {\bibfnamefont {T.~L.}\ \bibnamefont
  {Nicholson}}, \bibinfo {author} {\bibfnamefont {J.~R.}\ \bibnamefont
  {Williams}}, \bibinfo {author} {\bibfnamefont {S.~L.}\ \bibnamefont
  {Campbell}}, \bibinfo {author} {\bibfnamefont {M.}~\bibnamefont {Bishof}},
  \bibinfo {author} {\bibfnamefont {X.}~\bibnamefont {Zhang}}, \bibinfo
  {author} {\bibfnamefont {W.}~\bibnamefont {Zhang}}, \bibinfo {author}
  {\bibfnamefont {S.~L.}\ \bibnamefont {Bromley}}, \ and\ \bibinfo {author}
  {\bibfnamefont {J.}~\bibnamefont {Ye}},\ }\href {\doibase
  10.1038/nature12941} {\bibfield  {journal} {\bibinfo  {journal} {Nature}\
  }\textbf {\bibinfo {volume} {506}},\ \bibinfo {pages} {71} (\bibinfo {year}
  {2014})}\BibitemShut {NoStop}%
\bibitem [{\citenamefont {Hinkley}\ \emph {et~al.}(2013)\citenamefont
  {Hinkley}, \citenamefont {Sherman}, \citenamefont {Phillips}, \citenamefont
  {Schioppo}, \citenamefont {Lemke}, \citenamefont {Beloy}, \citenamefont
  {Pizzocaro}, \citenamefont {Oates},\ and\ \citenamefont
  {Ludlow}}]{Hinkley2013}%
  \BibitemOpen
  \bibfield  {author} {\bibinfo {author} {\bibfnamefont {N.}~\bibnamefont
  {Hinkley}}, \bibinfo {author} {\bibfnamefont {J.~A.}\ \bibnamefont
  {Sherman}}, \bibinfo {author} {\bibfnamefont {N.~B.}\ \bibnamefont
  {Phillips}}, \bibinfo {author} {\bibfnamefont {M.}~\bibnamefont {Schioppo}},
  \bibinfo {author} {\bibfnamefont {N.~D.}\ \bibnamefont {Lemke}}, \bibinfo
  {author} {\bibfnamefont {K.}~\bibnamefont {Beloy}}, \bibinfo {author}
  {\bibfnamefont {M.}~\bibnamefont {Pizzocaro}}, \bibinfo {author}
  {\bibfnamefont {C.~W.}\ \bibnamefont {Oates}}, \ and\ \bibinfo {author}
  {\bibfnamefont {A.~D.}\ \bibnamefont {Ludlow}},\ }\href {\doibase
  10.1126/science.1240420} {\bibfield  {journal} {\bibinfo  {journal} {Science
  (New York, N.Y.)}\ }\textbf {\bibinfo {volume} {341}},\ \bibinfo {pages}
  {1215} (\bibinfo {year} {2013})}\BibitemShut {NoStop}%
\bibitem [{\citenamefont {Beloy}\ \emph {et~al.}(2014)\citenamefont {Beloy},
  \citenamefont {Hinkley}, \citenamefont {Phillips}, \citenamefont {Sherman},
  \citenamefont {Schioppo}, \citenamefont {Lehman}, \citenamefont {Feldman},
  \citenamefont {Hanssen}, \citenamefont {Oates},\ and\ \citenamefont
  {Ludlow}}]{Beloy2014}%
  \BibitemOpen
  \bibfield  {author} {\bibinfo {author} {\bibfnamefont {K.}~\bibnamefont
  {Beloy}}, \bibinfo {author} {\bibfnamefont {N.}~\bibnamefont {Hinkley}},
  \bibinfo {author} {\bibfnamefont {N.~B.}\ \bibnamefont {Phillips}}, \bibinfo
  {author} {\bibfnamefont {J.~A.}\ \bibnamefont {Sherman}}, \bibinfo {author}
  {\bibfnamefont {M.}~\bibnamefont {Schioppo}}, \bibinfo {author}
  {\bibfnamefont {J.}~\bibnamefont {Lehman}}, \bibinfo {author} {\bibfnamefont
  {A.}~\bibnamefont {Feldman}}, \bibinfo {author} {\bibfnamefont {L.~M.}\
  \bibnamefont {Hanssen}}, \bibinfo {author} {\bibfnamefont {C.~W.}\
  \bibnamefont {Oates}}, \ and\ \bibinfo {author} {\bibfnamefont {A.~D.}\
  \bibnamefont {Ludlow}},\ }\href {\doibase 10.1103/PhysRevLett.113.260801}
  {\bibfield  {journal} {\bibinfo  {journal} {Physical Review Letters}\
  }\textbf {\bibinfo {volume} {113}},\ \bibinfo {pages} {260801} (\bibinfo
  {year} {2014})}\BibitemShut {NoStop}%
\bibitem [{\citenamefont {Safronova}\ \emph {et~al.}(2013)\citenamefont
  {Safronova}, \citenamefont {Porsev}, \citenamefont {Safronova}, \citenamefont
  {Kozlov},\ and\ \citenamefont {Clark}}]{Safronova2013}%
  \BibitemOpen
  \bibfield  {author} {\bibinfo {author} {\bibfnamefont {M.~S.}\ \bibnamefont
  {Safronova}}, \bibinfo {author} {\bibfnamefont {S.~G.}\ \bibnamefont
  {Porsev}}, \bibinfo {author} {\bibfnamefont {U.~I.}\ \bibnamefont
  {Safronova}}, \bibinfo {author} {\bibfnamefont {M.~G.}\ \bibnamefont
  {Kozlov}}, \ and\ \bibinfo {author} {\bibfnamefont {C.~W.}\ \bibnamefont
  {Clark}},\ }\href {\doibase 10.1103/PhysRevA.87.012509} {\bibfield  {journal}
  {\bibinfo  {journal} {Physical Review A}\ }\textbf {\bibinfo {volume} {87}},\
  \bibinfo {pages} {012509} (\bibinfo {year} {2013})}\BibitemShut {NoStop}%
\bibitem [{\citenamefont {Ushijima}\ \emph {et~al.}(2015)\citenamefont
  {Ushijima}, \citenamefont {Takamoto}, \citenamefont {Das}, \citenamefont
  {Ohkubo},\ and\ \citenamefont {Katori}}]{Ushijima2015}%
  \BibitemOpen
  \bibfield  {author} {\bibinfo {author} {\bibfnamefont {I.}~\bibnamefont
  {Ushijima}}, \bibinfo {author} {\bibfnamefont {M.}~\bibnamefont {Takamoto}},
  \bibinfo {author} {\bibfnamefont {M.}~\bibnamefont {Das}}, \bibinfo {author}
  {\bibfnamefont {T.}~\bibnamefont {Ohkubo}}, \ and\ \bibinfo {author}
  {\bibfnamefont {H.}~\bibnamefont {Katori}},\ }\href {\doibase
  10.1038/nphoton.2015.5} {\bibfield  {journal} {\bibinfo  {journal} {Nature
  Photonics}\ }\textbf {\bibinfo {volume} {9}},\ \bibinfo {pages} {185}
  (\bibinfo {year} {2015})}\BibitemShut {NoStop}%
\bibitem [{\citenamefont {Kulosa}\ \emph {et~al.}(2015)\citenamefont {Kulosa},
  \citenamefont {Fim}, \citenamefont {Zipfel}, \citenamefont {R{\"{u}}hmann},
  \citenamefont {Sauer}, \citenamefont {Jha}, \citenamefont {Gibble},
  \citenamefont {Ertmer}, \citenamefont {Rasel}, \citenamefont {Safronova},
  \citenamefont {Safronova},\ and\ \citenamefont {Porsev}}]{Kulosa2015}%
  \BibitemOpen
  \bibfield  {author} {\bibinfo {author} {\bibfnamefont {A.~P.}\ \bibnamefont
  {Kulosa}}, \bibinfo {author} {\bibfnamefont {D.}~\bibnamefont {Fim}},
  \bibinfo {author} {\bibfnamefont {K.~H.}\ \bibnamefont {Zipfel}}, \bibinfo
  {author} {\bibfnamefont {S.}~\bibnamefont {R{\"{u}}hmann}}, \bibinfo {author}
  {\bibfnamefont {S.}~\bibnamefont {Sauer}}, \bibinfo {author} {\bibfnamefont
  {N.}~\bibnamefont {Jha}}, \bibinfo {author} {\bibfnamefont {K.}~\bibnamefont
  {Gibble}}, \bibinfo {author} {\bibfnamefont {W.}~\bibnamefont {Ertmer}},
  \bibinfo {author} {\bibfnamefont {E.~M.}\ \bibnamefont {Rasel}}, \bibinfo
  {author} {\bibfnamefont {M.~S.}\ \bibnamefont {Safronova}}, \bibinfo {author}
  {\bibfnamefont {U.~I.}\ \bibnamefont {Safronova}}, \ and\ \bibinfo {author}
  {\bibfnamefont {S.~G.}\ \bibnamefont {Porsev}},\ }\href {\doibase
  10.1103/PhysRevLett.115.240801} {\bibfield  {journal} {\bibinfo  {journal}
  {Physical Review Letters}\ }\textbf {\bibinfo {volume} {115}},\ \bibinfo
  {pages} {240801} (\bibinfo {year} {2015})},\ \Eprint
  {http://arxiv.org/abs/1508.01118} {arXiv:1508.01118} \BibitemShut {NoStop}%
\bibitem [{\citenamefont {McFerran}\ \emph {et~al.}(2014)\citenamefont
  {McFerran}, \citenamefont {Yi}, \citenamefont {Mejri}, \citenamefont {Zhang},
  \citenamefont {{Di Manno}}, \citenamefont {Abgrall}, \citenamefont
  {Gu{\'{e}}na}, \citenamefont {{Le Coq}},\ and\ \citenamefont
  {Bize}}]{McFerran2014}%
  \BibitemOpen
  \bibfield  {author} {\bibinfo {author} {\bibfnamefont {J.~J.}\ \bibnamefont
  {McFerran}}, \bibinfo {author} {\bibfnamefont {L.}~\bibnamefont {Yi}},
  \bibinfo {author} {\bibfnamefont {S.}~\bibnamefont {Mejri}}, \bibinfo
  {author} {\bibfnamefont {W.}~\bibnamefont {Zhang}}, \bibinfo {author}
  {\bibfnamefont {S.}~\bibnamefont {{Di Manno}}}, \bibinfo {author}
  {\bibfnamefont {M.}~\bibnamefont {Abgrall}}, \bibinfo {author} {\bibfnamefont
  {J.}~\bibnamefont {Gu{\'{e}}na}}, \bibinfo {author} {\bibfnamefont
  {Y.}~\bibnamefont {{Le Coq}}}, \ and\ \bibinfo {author} {\bibfnamefont
  {S.}~\bibnamefont {Bize}},\ }\href {\doibase 10.1103/PhysRevA.89.043432}
  {\bibfield  {journal} {\bibinfo  {journal} {Physical Review A}\ }\textbf
  {\bibinfo {volume} {89}},\ \bibinfo {pages} {043432} (\bibinfo {year}
  {2014})}\BibitemShut {NoStop}%
\bibitem [{\citenamefont {Aikawa}\ \emph {et~al.}(2012)\citenamefont {Aikawa},
  \citenamefont {Frisch}, \citenamefont {Mark}, \citenamefont {Baier},
  \citenamefont {Rietzler}, \citenamefont {Grimm},\ and\ \citenamefont
  {Ferlaino}}]{Aikawa2012}%
  \BibitemOpen
  \bibfield  {author} {\bibinfo {author} {\bibfnamefont {K.}~\bibnamefont
  {Aikawa}}, \bibinfo {author} {\bibfnamefont {A.}~\bibnamefont {Frisch}},
  \bibinfo {author} {\bibfnamefont {M.}~\bibnamefont {Mark}}, \bibinfo {author}
  {\bibfnamefont {S.}~\bibnamefont {Baier}}, \bibinfo {author} {\bibfnamefont
  {A.}~\bibnamefont {Rietzler}}, \bibinfo {author} {\bibfnamefont
  {R.}~\bibnamefont {Grimm}}, \ and\ \bibinfo {author} {\bibfnamefont
  {F.}~\bibnamefont {Ferlaino}},\ }\href {\doibase
  10.1103/PhysRevLett.108.210401} {\bibfield  {journal} {\bibinfo  {journal}
  {Physical Review Letters}\ }\textbf {\bibinfo {volume} {108}},\ \bibinfo
  {pages} {210401} (\bibinfo {year} {2012})}\BibitemShut {NoStop}%
\bibitem [{\citenamefont {Lu}\ \emph {et~al.}(2011)\citenamefont {Lu},
  \citenamefont {Burdick}, \citenamefont {Youn},\ and\ \citenamefont
  {Lev}}]{ref:DyBEC:Lev:PRL}%
  \BibitemOpen
  \bibfield  {author} {\bibinfo {author} {\bibfnamefont {M.}~\bibnamefont
  {Lu}}, \bibinfo {author} {\bibfnamefont {N.~Q.}\ \bibnamefont {Burdick}},
  \bibinfo {author} {\bibfnamefont {S.~H.}\ \bibnamefont {Youn}}, \ and\
  \bibinfo {author} {\bibfnamefont {B.~L.}\ \bibnamefont {Lev}},\ }\href
  {\doibase 10.1103/PhysRevLett.107.190401} {\bibfield  {journal} {\bibinfo
  {journal} {Phys. Rev. Lett.}\ }\textbf {\bibinfo {volume} {107}},\ \bibinfo
  {pages} {190401} (\bibinfo {year} {2011})}\BibitemShut {NoStop}%
\bibitem [{\citenamefont {Sukachev}\ \emph {et~al.}(2010)\citenamefont
  {Sukachev}, \citenamefont {Sokolov}, \citenamefont {Chebakov}, \citenamefont
  {Akimov}, \citenamefont {Kanorsky}, \citenamefont {Kolachevsky},\ and\
  \citenamefont {Sorokin}}]{Sukachev2010}%
  \BibitemOpen
  \bibfield  {author} {\bibinfo {author} {\bibfnamefont {D.}~\bibnamefont
  {Sukachev}}, \bibinfo {author} {\bibfnamefont {A.}~\bibnamefont {Sokolov}},
  \bibinfo {author} {\bibfnamefont {K.}~\bibnamefont {Chebakov}}, \bibinfo
  {author} {\bibfnamefont {A.}~\bibnamefont {Akimov}}, \bibinfo {author}
  {\bibfnamefont {S.}~\bibnamefont {Kanorsky}}, \bibinfo {author}
  {\bibfnamefont {N.}~\bibnamefont {Kolachevsky}}, \ and\ \bibinfo {author}
  {\bibfnamefont {V.}~\bibnamefont {Sorokin}},\ }\href@noop {} {\bibfield
  {journal} {\bibinfo  {journal} {Physical Review A}\ }\textbf {\bibinfo
  {volume} {82}} (\bibinfo {year} {2010})}\BibitemShut {NoStop}%
\bibitem [{\citenamefont {Sukachev}\ \emph {et~al.}(2014)\citenamefont
  {Sukachev}, \citenamefont {Kalganova}, \citenamefont {Sokolov}, \citenamefont
  {Fedorov}, \citenamefont {Vishnyakova}, \citenamefont {Akimov}, \citenamefont
  {Kolachevsky},\ and\ \citenamefont {Sorokin}}]{Sukachev2014}%
  \BibitemOpen
  \bibfield  {author} {\bibinfo {author} {\bibfnamefont {D.~D.}\ \bibnamefont
  {Sukachev}}, \bibinfo {author} {\bibfnamefont {E.~S.}\ \bibnamefont
  {Kalganova}}, \bibinfo {author} {\bibfnamefont {A.~V.}\ \bibnamefont
  {Sokolov}}, \bibinfo {author} {\bibfnamefont {S.~A.}\ \bibnamefont
  {Fedorov}}, \bibinfo {author} {\bibfnamefont {G.~A.}\ \bibnamefont
  {Vishnyakova}}, \bibinfo {author} {\bibfnamefont {A.~V.}\ \bibnamefont
  {Akimov}}, \bibinfo {author} {\bibfnamefont {N.~N.}\ \bibnamefont
  {Kolachevsky}}, \ and\ \bibinfo {author} {\bibfnamefont {V.~N.}\ \bibnamefont
  {Sorokin}},\ }\href {\doibase 10.1070/QE2014v044n06ABEH015392} {\bibfield
  {journal} {\bibinfo  {journal} {Quantum Electronics}\ }\textbf {\bibinfo
  {volume} {44}},\ \bibinfo {pages} {515} (\bibinfo {year} {2014})}\BibitemShut
  {NoStop}%
\bibitem [{\citenamefont {Alnis}\ \emph {et~al.}(2008)\citenamefont {Alnis},
  \citenamefont {Matveev}, \citenamefont {Kolachevsky}, \citenamefont {Udem},\
  and\ \citenamefont {H{\"{a}}nsch}}]{Alnis2008}%
  \BibitemOpen
  \bibfield  {author} {\bibinfo {author} {\bibfnamefont {J.}~\bibnamefont
  {Alnis}}, \bibinfo {author} {\bibfnamefont {A.}~\bibnamefont {Matveev}},
  \bibinfo {author} {\bibfnamefont {N.}~\bibnamefont {Kolachevsky}}, \bibinfo
  {author} {\bibfnamefont {T.}~\bibnamefont {Udem}}, \ and\ \bibinfo {author}
  {\bibfnamefont {T.~W.}\ \bibnamefont {H{\"{a}}nsch}},\ }\href {\doibase
  10.1103/PhysRevA.77.053809} {\bibfield  {journal} {\bibinfo  {journal}
  {Physical Review A}\ }\textbf {\bibinfo {volume} {77}},\ \bibinfo {pages}
  {053809} (\bibinfo {year} {2008})}\BibitemShut {NoStop}%
\bibitem [{\citenamefont {Kessler}\ \emph {et~al.}(2012)\citenamefont
  {Kessler}, \citenamefont {Hagemann}, \citenamefont {Grebing}, \citenamefont
  {Legero}, \citenamefont {Sterr}, \citenamefont {Riehle}, \citenamefont
  {Martin}, \citenamefont {Chen},\ and\ \citenamefont {Ye}}]{Kessler2012}%
  \BibitemOpen
  \bibfield  {author} {\bibinfo {author} {\bibfnamefont {T.}~\bibnamefont
  {Kessler}}, \bibinfo {author} {\bibfnamefont {C.}~\bibnamefont {Hagemann}},
  \bibinfo {author} {\bibfnamefont {C.}~\bibnamefont {Grebing}}, \bibinfo
  {author} {\bibfnamefont {T.}~\bibnamefont {Legero}}, \bibinfo {author}
  {\bibfnamefont {U.}~\bibnamefont {Sterr}}, \bibinfo {author} {\bibfnamefont
  {F.}~\bibnamefont {Riehle}}, \bibinfo {author} {\bibfnamefont {M.~J.}\
  \bibnamefont {Martin}}, \bibinfo {author} {\bibfnamefont {L.}~\bibnamefont
  {Chen}}, \ and\ \bibinfo {author} {\bibfnamefont {J.}~\bibnamefont {Ye}},\
  }\href {\doibase 10.1038/nphoton.2012.217} {\bibfield  {journal} {\bibinfo
  {journal} {Nature Photonics}\ }\textbf {\bibinfo {volume} {6}},\ \bibinfo
  {pages} {687} (\bibinfo {year} {2012})}\BibitemShut {NoStop}%
\bibitem [{\citenamefont {{Vishnyakova, G A, Golovizin, A A Kalganova}}\ \emph
  {et~al.}(2016)\citenamefont {{Vishnyakova, G A, Golovizin, A A Kalganova}},
  \citenamefont {Sorokin},\ and\ \citenamefont {{Sukachev, D D Tregubov, D O
  Khabarova, K Yu Kolachevsky}}}]{Vishnyakova2016}%
  \BibitemOpen
  \bibfield  {author} {\bibinfo {author} {\bibfnamefont {E.~K.}\ \bibnamefont
  {{Vishnyakova, G A, Golovizin, A A Kalganova}}}, \bibinfo {author}
  {\bibfnamefont {V.~N.}\ \bibnamefont {Sorokin}}, \ and\ \bibinfo {author}
  {\bibfnamefont {N.~N.}\ \bibnamefont {{Sukachev, D D Tregubov, D O Khabarova,
  K Yu Kolachevsky}}},\ }\href
  {http://iopscience.iop.org.ezp-prod1.hul.harvard.edu/article/10.3367/UFNe.0186.201602h.0176/meta}
  {\bibfield  {journal} {\bibinfo  {journal} {Physics-Uspekhi}\ }\textbf
  {\bibinfo {volume} {59}},\ \bibinfo {pages} {168} (\bibinfo {year}
  {2016})}\BibitemShut {NoStop}%
\bibitem [{\citenamefont {Frisch}(2014)}]{Frish-PhD-2014}%
  \BibitemOpen
  \bibfield  {author} {\bibinfo {author} {\bibfnamefont {A.}~\bibnamefont
  {Frisch}},\ }\emph {\bibinfo {title} {{Dipolar Quantum Gases of Erbium}}},\
  \href
  {http://www.ultracold.at/theses/thesis_albert_frisch/thesis_albert_frisch.pdf}
  {Ph.D. thesis},\ \bibinfo  {school} {University of Inssbruck} (\bibinfo
  {year} {2014})\BibitemShut {NoStop}%
\bibitem [{\citenamefont {Ketterle}\ and\ \citenamefont {{Van
  Druten}}(1996)}]{Ketterle1996}%
  \BibitemOpen
  \bibfield  {author} {\bibinfo {author} {\bibfnamefont {W.}~\bibnamefont
  {Ketterle}}\ and\ \bibinfo {author} {\bibfnamefont {N.~J.}\ \bibnamefont
  {{Van Druten}}},\ }in\ \href
  {http://www.sciencedirect.com/science/bookseries/1049250X/37} {\emph
  {\bibinfo {booktitle} {Atomic, Molecular, and Optical Physics Volume 37}}},\
  \bibinfo {editor} {edited by\ \bibinfo {editor} {\bibfnamefont {B.~B.}\
  \bibnamefont {Walther}}\ and\ \bibinfo {editor} {\bibnamefont {Herbert}}}\
  (\bibinfo  {publisher} {Academic Press},\ \bibinfo {year} {1996})\ pp.\
  \bibinfo {pages} {181--236}\BibitemShut {NoStop}%
\bibitem [{\citenamefont {Dalibard}(1998)}]{Dalibard1998}%
  \BibitemOpen
  \bibfield  {author} {\bibinfo {author} {\bibfnamefont {J.}~\bibnamefont
  {Dalibard}},\ }\href
  {https://books.google.com/books?id=qZPCAQAAQBAJ&pg=PA321&lpg=PA321&dq=Collisional+dynamics+of+ultra-cold+atomic+gases.+Dalibard&source=bl&ots=wxL5rHDxOb&sig=kC_rpxpNquk5szPfoKNZ4pYNwGU&hl=ru&sa=X&ved=0ahUKEwius6eXzP_MAhVLOD4KHZ4qDWoQ6AEISzAF#v=onepage&q=Collisional
  dynamics of ultra-cold atomic gases. Dalibard&f=false} {\emph {\bibinfo
  {title} {Proceedings of the International School of Physics ’Enrico
  Fermi’, Course CXL: ’Bose- Einstein condensation in gases’}}},\ edited
  by\ \bibinfo {editor} {\bibfnamefont {M.}~\bibnamefont {Inguscio}}, \bibinfo
  {editor} {\bibfnamefont {S.}~\bibnamefont {Stringari}}, \ and\ \bibinfo
  {editor} {\bibfnamefont {C.}~\bibnamefont {Wieman}}\ (\bibinfo  {publisher}
  {IOS press},\ \bibinfo {address} {Varenna},\ \bibinfo {year} {1998})\ pp.\
  \bibinfo {pages} {321--350}\BibitemShut {NoStop}%
\bibitem [{\citenamefont {Gribakin}\ and\ \citenamefont
  {Flambaum}(1993)}]{Gribakin1993}%
  \BibitemOpen
  \bibfield  {author} {\bibinfo {author} {\bibfnamefont {G.~F.}\ \bibnamefont
  {Gribakin}}\ and\ \bibinfo {author} {\bibfnamefont {V.~V.}\ \bibnamefont
  {Flambaum}},\ }\href {\doibase 10.1103/PhysRevA.48.546} {\bibfield  {journal}
  {\bibinfo  {journal} {Physical Review A}\ }\textbf {\bibinfo {volume} {48}},\
  \bibinfo {pages} {546} (\bibinfo {year} {1993})}\BibitemShut {NoStop}%
\bibitem [{\citenamefont {Golovizin}\ \emph {et~al.}(2015)\citenamefont
  {Golovizin}, \citenamefont {Kalganova}, \citenamefont {Sukachev},
  \citenamefont {Vishnyakova}, \citenamefont {Semerikov}, \citenamefont
  {Soshenko}, \citenamefont {Tregubov}, \citenamefont {Akimov}, \citenamefont
  {Kolachevsky}, \citenamefont {Khabarova},\ and\ \citenamefont
  {Sorokin}}]{Golovizin2015}%
  \BibitemOpen
  \bibfield  {author} {\bibinfo {author} {\bibfnamefont {A.~A.}\ \bibnamefont
  {Golovizin}}, \bibinfo {author} {\bibfnamefont {E.~S.}\ \bibnamefont
  {Kalganova}}, \bibinfo {author} {\bibfnamefont {D.~D.}\ \bibnamefont
  {Sukachev}}, \bibinfo {author} {\bibfnamefont {G.~A.}\ \bibnamefont
  {Vishnyakova}}, \bibinfo {author} {\bibfnamefont {I.~A.}\ \bibnamefont
  {Semerikov}}, \bibinfo {author} {\bibfnamefont {V.~V.}\ \bibnamefont
  {Soshenko}}, \bibinfo {author} {\bibfnamefont {D.~O.}\ \bibnamefont
  {Tregubov}}, \bibinfo {author} {\bibfnamefont {A.~V.}\ \bibnamefont
  {Akimov}}, \bibinfo {author} {\bibfnamefont {N.~N.}\ \bibnamefont
  {Kolachevsky}}, \bibinfo {author} {\bibfnamefont {K.~Y.}\ \bibnamefont
  {Khabarova}}, \ and\ \bibinfo {author} {\bibfnamefont {V.~N.}\ \bibnamefont
  {Sorokin}},\ }\href {\doibase 10.1070/QE2015v045n05ABEH015749} {\bibfield
  {journal} {\bibinfo  {journal} {Quantum Electronics}\ }\textbf {\bibinfo
  {volume} {45}},\ \bibinfo {pages} {482} (\bibinfo {year} {2015})}\BibitemShut
  {NoStop}%
\bibitem [{\citenamefont {Rinkleff}\ and\ \citenamefont
  {Thorn}(1994)}]{RETensorPol}%
  \BibitemOpen
  \bibfield  {author} {\bibinfo {author} {\bibfnamefont {R.-H.}\ \bibnamefont
  {Rinkleff}}\ and\ \bibinfo {author} {\bibfnamefont {F.}~\bibnamefont
  {Thorn}},\ }\href {\doibase 10.1007/BF01437144} {\bibfield  {journal}
  {\bibinfo  {journal} {Zeitschrift f{\"{u}}r Physik D Atoms, Molecules and
  Clusters}\ }\textbf {\bibinfo {volume} {32}},\ \bibinfo {pages} {173}
  (\bibinfo {year} {1994})}\BibitemShut {NoStop}%
\bibitem [{\citenamefont {Rinkleff}(1978)}]{TmTensorPol}%
  \BibitemOpen
  \bibfield  {author} {\bibinfo {author} {\bibfnamefont {R.-H.}\ \bibnamefont
  {Rinkleff}},\ }\href {\doibase 10.1007/BF01415190} {\bibfield  {journal}
  {\bibinfo  {journal} {Zeitschrift f{\"{u}}r Physik A Atoms and Nuclei}\
  }\textbf {\bibinfo {volume} {288}},\ \bibinfo {pages} {233} (\bibinfo {year}
  {1978})}\BibitemShut {NoStop}%
\bibitem [{\citenamefont {Chu}\ \emph {et~al.}(2007)\citenamefont {Chu},
  \citenamefont {Dalgarno},\ and\ \citenamefont {Groenenboom}}]{Chu2007}%
  \BibitemOpen
  \bibfield  {author} {\bibinfo {author} {\bibfnamefont {X.}~\bibnamefont
  {Chu}}, \bibinfo {author} {\bibfnamefont {A.}~\bibnamefont {Dalgarno}}, \
  and\ \bibinfo {author} {\bibfnamefont {G.~C.}\ \bibnamefont {Groenenboom}},\
  }\href {\doibase 10.1103/PhysRevA.75.032723} {\bibfield  {journal} {\bibinfo
  {journal} {Phys. Rev. A}\ }\textbf {\bibinfo {volume} {75}},\ \bibinfo
  {pages} {032723} (\bibinfo {year} {2007})}\BibitemShut {NoStop}%
\bibitem [{\citenamefont {Lepers}\ \emph {et~al.}(2014)\citenamefont {Lepers},
  \citenamefont {Wyart},\ and\ \citenamefont {Dulieu}}]{Lepers2014}%
  \BibitemOpen
  \bibfield  {author} {\bibinfo {author} {\bibfnamefont {M.}~\bibnamefont
  {Lepers}}, \bibinfo {author} {\bibfnamefont {J.-F.}\ \bibnamefont {Wyart}}, \
  and\ \bibinfo {author} {\bibfnamefont {O.}~\bibnamefont {Dulieu}},\ }\href
  {\doibase 10.1103/PhysRevA.89.022505} {\bibfield  {journal} {\bibinfo
  {journal} {Physical Review A}\ }\textbf {\bibinfo {volume} {89}},\ \bibinfo
  {pages} {022505} (\bibinfo {year} {2014})}\BibitemShut {NoStop}%
\bibitem [{\citenamefont {Mitroy}\ \emph {et~al.}(2010)\citenamefont {Mitroy},
  \citenamefont {Safronova},\ and\ \citenamefont {Clark}}]{Mitroy2010}%
  \BibitemOpen
  \bibfield  {author} {\bibinfo {author} {\bibfnamefont {J.}~\bibnamefont
  {Mitroy}}, \bibinfo {author} {\bibfnamefont {M.~S.}\ \bibnamefont
  {Safronova}}, \ and\ \bibinfo {author} {\bibfnamefont {C.~W.}\ \bibnamefont
  {Clark}},\ }\href {\doibase 10.1088/0953-4075/43/20/202001} {\bibfield
  {journal} {\bibinfo  {journal} {Journal of Physics B: Atomic, Molecular and
  Optical Physics}\ }\textbf {\bibinfo {volume} {43}},\ \bibinfo {pages}
  {202001} (\bibinfo {year} {2010})}\BibitemShut {NoStop}%
\bibitem [{\citenamefont {Angel}\ and\ \citenamefont
  {Sandars}(1968)}]{Angel1968}%
  \BibitemOpen
  \bibfield  {author} {\bibinfo {author} {\bibfnamefont {J.~R.~P.}\
  \bibnamefont {Angel}}\ and\ \bibinfo {author} {\bibfnamefont {P.~G.~H.}\
  \bibnamefont {Sandars}},\ }\href {\doibase 10.1098/rspa.1968.0109} {\bibfield
   {journal} {\bibinfo  {journal} {Proceedings of the Royal Society A:
  Mathematical, Physical and Engineering Sciences}\ }\textbf {\bibinfo {volume}
  {305}},\ \bibinfo {pages} {125} (\bibinfo {year} {1968})}\BibitemShut
  {NoStop}%
\bibitem [{\citenamefont {Wickliffe}\ and\ \citenamefont
  {Lawler}(1997)}]{TmTransitions}%
  \BibitemOpen
  \bibfield  {author} {\bibinfo {author} {\bibfnamefont {M.~E.}\ \bibnamefont
  {Wickliffe}}\ and\ \bibinfo {author} {\bibfnamefont {J.~E.}\ \bibnamefont
  {Lawler}},\ }\href {\doibase 10.1364/JOSAB.14.000737} {\bibfield  {journal}
  {\bibinfo  {journal} {Journal of the Optical Society of America B}\ }\textbf
  {\bibinfo {volume} {14}},\ \bibinfo {pages} {737} (\bibinfo {year}
  {1997})}\BibitemShut {NoStop}%
\bibitem [{\citenamefont {Anderson}\ \emph {et~al.}(1996)\citenamefont
  {Anderson}, \citenamefont {Hartog},\ and\ \citenamefont
  {Lawler}}]{Anderson1996}%
  \BibitemOpen
  \bibfield  {author} {\bibinfo {author} {\bibfnamefont {H.~M.}\ \bibnamefont
  {Anderson}}, \bibinfo {author} {\bibfnamefont {E.~A.~D.}\ \bibnamefont
  {Hartog}}, \ and\ \bibinfo {author} {\bibfnamefont {J.~E.}\ \bibnamefont
  {Lawler}},\ }\href {\doibase 10.1364/JOSAB.13.002382} {\bibfield  {journal}
  {\bibinfo  {journal} {Journal of the Optical Society of America B}\ }\textbf
  {\bibinfo {volume} {13}},\ \bibinfo {pages} {2382} (\bibinfo {year}
  {1996})}\BibitemShut {NoStop}%
\bibitem [{\citenamefont {Cowan}()}]{Cowan1981}%
  \BibitemOpen
  \bibfield  {author} {\bibinfo {author} {\bibfnamefont {R.}~\bibnamefont
  {Cowan}},\ }\href@noop {} {\emph {\bibinfo {title} {{The Theory of Atomic
  Structure and Spectra (University of California Press, Berkeley, CA, 1981),
  and Cowan programs RCN, RCN2, and RCG}}}}\BibitemShut {NoStop}%
\bibitem [{\citenamefont {Veseth}\ and\ \citenamefont
  {Kelly}(1992)}]{Veseth1992}%
  \BibitemOpen
  \bibfield  {author} {\bibinfo {author} {\bibfnamefont {L.}~\bibnamefont
  {Veseth}}\ and\ \bibinfo {author} {\bibfnamefont {H.~P.}\ \bibnamefont
  {Kelly}},\ }\href {\doibase 10.1103/PhysRevA.45.4621} {\bibfield  {journal}
  {\bibinfo  {journal} {Physical Review A}\ }\textbf {\bibinfo {volume} {45}},\
  \bibinfo {pages} {4621} (\bibinfo {year} {1992})}\BibitemShut {NoStop}%
\bibitem [{\citenamefont {Gu}(2008)}]{FAC2011}%
  \BibitemOpen
  \bibfield  {author} {\bibinfo {author} {\bibfnamefont {M.~F.}\ \bibnamefont
  {Gu}},\ }\href
  {http://www.nrcresearchpress.com/doi/abs/10.1139/p07-197?journalCode=cjp&#.VovWO_nWE2Y}
  {\bibfield  {journal} {\bibinfo  {journal} {Canadian Journal of Physics}\
  }\textbf {\bibinfo {volume} {86}},\ \bibinfo {pages} {675} (\bibinfo {year}
  {2008})}\BibitemShut {NoStop}%
\bibitem [{\citenamefont {Whitfield}\ \emph {et~al.}(2008)\citenamefont
  {Whitfield}, \citenamefont {Caspary}, \citenamefont {Wehlitz},\ and\
  \citenamefont {Martins}}]{ref:Whitfield2008}%
  \BibitemOpen
  \bibfield  {author} {\bibinfo {author} {\bibfnamefont {S.~B.}\ \bibnamefont
  {Whitfield}}, \bibinfo {author} {\bibfnamefont {K.}~\bibnamefont {Caspary}},
  \bibinfo {author} {\bibfnamefont {R.}~\bibnamefont {Wehlitz}}, \ and\
  \bibinfo {author} {\bibfnamefont {M.}~\bibnamefont {Martins}},\ }\href
  {\doibase 10.1088/0953-4075/41/1/015001} {\bibfield  {journal} {\bibinfo
  {journal} {Journal of Physics B}\ }\textbf {\bibinfo {volume} {41}},\
  \bibinfo {pages} {015001} (\bibinfo {year} {2008})}\BibitemShut {NoStop}%
\bibitem [{\citenamefont {Bishop}(1994)}]{Bishop1994}%
  \BibitemOpen
  \bibfield  {author} {\bibinfo {author} {\bibfnamefont {D.~M.}\ \bibnamefont
  {Bishop}},\ }\href {\doibase 10.1063/1.467062} {\bibfield  {journal}
  {\bibinfo  {journal} {The Journal of Chemical Physics}\ }\textbf {\bibinfo
  {volume} {100}},\ \bibinfo {pages} {6535} (\bibinfo {year}
  {1994})}\BibitemShut {NoStop}%
\bibitem [{\citenamefont {Giglberger}\ and\ \citenamefont
  {Penselin}(1967)}]{Giglberger1967}%
  \BibitemOpen
  \bibfield  {author} {\bibinfo {author} {\bibfnamefont {D.}~\bibnamefont
  {Giglberger}}\ and\ \bibinfo {author} {\bibfnamefont {S.}~\bibnamefont
  {Penselin}},\ }\href {\doibase 10.1007/BF01326434} {\bibfield  {journal}
  {\bibinfo  {journal} {Zeitschrift fuer Physik}\ }\textbf {\bibinfo {volume}
  {199}},\ \bibinfo {pages} {244} (\bibinfo {year} {1967})}\BibitemShut
  {NoStop}%
\bibitem [{\citenamefont {van Leeuwen}\ \emph {et~al.}(1980)\citenamefont {van
  Leeuwen}, \citenamefont {Eliel},\ and\ \citenamefont
  {Hogervorst}}]{Leeuwen1980}%
  \BibitemOpen
  \bibfield  {author} {\bibinfo {author} {\bibfnamefont {K.}~\bibnamefont {van
  Leeuwen}}, \bibinfo {author} {\bibfnamefont {E.}~\bibnamefont {Eliel}}, \
  and\ \bibinfo {author} {\bibfnamefont {W.}~\bibnamefont {Hogervorst}},\
  }\href {\doibase 10.1016/0375-9601(80)90805-1} {\bibfield  {journal}
  {\bibinfo  {journal} {Physics Letters A}\ }\textbf {\bibinfo {volume} {78}},\
  \bibinfo {pages} {54} (\bibinfo {year} {1980})}\BibitemShut {NoStop}%
\bibitem [{\citenamefont {Blaise}\ and\ \citenamefont
  {Pierre}(1965)}]{Blaise1965}%
  \BibitemOpen
  \bibfield  {author} {\bibinfo {author} {\bibfnamefont {J.}~\bibnamefont
  {Blaise}}\ and\ \bibinfo {author} {\bibfnamefont {C.}~\bibnamefont
  {Pierre}},\ }\href
  {https://archive.org/stream/ComptesRendusAcademieDesSciences0260/ComptesRendusAcadmieDesSciences-Tome260-Janvier-juin1965Partie5_djvu.txt}
  {\bibfield  {journal} {\bibinfo  {journal} {Comptes rendus de
  l'Acad{\'{e}}mie des sciences.}\ }\textbf {\bibinfo {volume} {260}},\
  \bibinfo {pages} {4693} (\bibinfo {year} {1965})}\BibitemShut {NoStop}%
\bibitem [{\citenamefont {Middelmann}\ \emph {et~al.}(2011)\citenamefont
  {Middelmann}, \citenamefont {Lisdat}, \citenamefont {Falke}, \citenamefont
  {Winfred}, \citenamefont {Riehle},\ and\ \citenamefont
  {Sterr}}]{middelmann2011}%
  \BibitemOpen
  \bibfield  {author} {\bibinfo {author} {\bibfnamefont {T.}~\bibnamefont
  {Middelmann}}, \bibinfo {author} {\bibfnamefont {C.}~\bibnamefont {Lisdat}},
  \bibinfo {author} {\bibfnamefont {S.}~\bibnamefont {Falke}}, \bibinfo
  {author} {\bibfnamefont {J.~S.~R.}\ \bibnamefont {Winfred}}, \bibinfo
  {author} {\bibfnamefont {F.}~\bibnamefont {Riehle}}, \ and\ \bibinfo {author}
  {\bibfnamefont {U.}~\bibnamefont {Sterr}},\ }\href@noop {} {\bibfield
  {journal} {\bibinfo  {journal} {IEEE Transactions on Instrumentation and
  Measurement}\ }\textbf {\bibinfo {volume} {60}},\ \bibinfo {pages} {2550}
  (\bibinfo {year} {2011})}\BibitemShut {NoStop}%
\bibitem [{\citenamefont {Rosenband}\ \emph {et~al.}(2007)\citenamefont
  {Rosenband}, \citenamefont {Schmidt}, \citenamefont {Hume}, \citenamefont
  {Itano}, \citenamefont {Fortier}, \citenamefont {Stalnaker}, \citenamefont
  {Kim}, \citenamefont {Diddams}, \citenamefont {Koelemeij}, \citenamefont
  {Bergquist},\ and\ \citenamefont {Wineland}}]{Rosenband2007}%
  \BibitemOpen
  \bibfield  {author} {\bibinfo {author} {\bibfnamefont {T.}~\bibnamefont
  {Rosenband}}, \bibinfo {author} {\bibfnamefont {P.}~\bibnamefont {Schmidt}},
  \bibinfo {author} {\bibfnamefont {D.}~\bibnamefont {Hume}}, \bibinfo {author}
  {\bibfnamefont {W.}~\bibnamefont {Itano}}, \bibinfo {author} {\bibfnamefont
  {T.}~\bibnamefont {Fortier}}, \bibinfo {author} {\bibfnamefont
  {J.}~\bibnamefont {Stalnaker}}, \bibinfo {author} {\bibfnamefont
  {K.}~\bibnamefont {Kim}}, \bibinfo {author} {\bibfnamefont {S.}~\bibnamefont
  {Diddams}}, \bibinfo {author} {\bibfnamefont {J.}~\bibnamefont {Koelemeij}},
  \bibinfo {author} {\bibfnamefont {J.}~\bibnamefont {Bergquist}}, \ and\
  \bibinfo {author} {\bibfnamefont {D.}~\bibnamefont {Wineland}},\ }\href
  {\doibase 10.1103/PhysRevLett.98.220801} {\bibfield  {journal} {\bibinfo
  {journal} {Physical Review Letters}\ }\textbf {\bibinfo {volume} {98}},\
  \bibinfo {pages} {220801} (\bibinfo {year} {2007})}\BibitemShut {NoStop}%
\bibitem [{\citenamefont {Rosenband}\ \emph {et~al.}(2008)\citenamefont
  {Rosenband}, \citenamefont {Hume}, \citenamefont {Schmidt}, \citenamefont
  {Chou}, \citenamefont {Brusch}, \citenamefont {Lorini}, \citenamefont
  {Oskay}, \citenamefont {Drullinger}, \citenamefont {Fortier}, \citenamefont
  {Stalnaker}, \citenamefont {Diddams}, \citenamefont {Swann}, \citenamefont
  {Newbury}, \citenamefont {Itano}, \citenamefont {Wineland},\ and\
  \citenamefont {Bergquist}}]{ref:IonOpticalClocks:NIST:Science}%
  \BibitemOpen
  \bibfield  {author} {\bibinfo {author} {\bibfnamefont {T.}~\bibnamefont
  {Rosenband}}, \bibinfo {author} {\bibfnamefont {D.~B.}\ \bibnamefont {Hume}},
  \bibinfo {author} {\bibfnamefont {P.~O.}\ \bibnamefont {Schmidt}}, \bibinfo
  {author} {\bibfnamefont {C.~W.}\ \bibnamefont {Chou}}, \bibinfo {author}
  {\bibfnamefont {A.}~\bibnamefont {Brusch}}, \bibinfo {author} {\bibfnamefont
  {L.}~\bibnamefont {Lorini}}, \bibinfo {author} {\bibfnamefont {W.~H.}\
  \bibnamefont {Oskay}}, \bibinfo {author} {\bibfnamefont {R.~E.}\ \bibnamefont
  {Drullinger}}, \bibinfo {author} {\bibfnamefont {T.~M.}\ \bibnamefont
  {Fortier}}, \bibinfo {author} {\bibfnamefont {J.~E.}\ \bibnamefont
  {Stalnaker}}, \bibinfo {author} {\bibfnamefont {S.~A.}\ \bibnamefont
  {Diddams}}, \bibinfo {author} {\bibfnamefont {W.~C.}\ \bibnamefont {Swann}},
  \bibinfo {author} {\bibfnamefont {N.~R.}\ \bibnamefont {Newbury}}, \bibinfo
  {author} {\bibfnamefont {W.~M.}\ \bibnamefont {Itano}}, \bibinfo {author}
  {\bibfnamefont {D.~J.}\ \bibnamefont {Wineland}}, \ and\ \bibinfo {author}
  {\bibfnamefont {J.~C.}\ \bibnamefont {Bergquist}},\ }\href {\doibase
  10.1126/science.1154622} {\bibfield  {journal} {\bibinfo  {journal}
  {Science}\ }\textbf {\bibinfo {volume} {319}},\ \bibinfo {pages} {1808}
  (\bibinfo {year} {2008})}\BibitemShut {NoStop}%
\bibitem [{\citenamefont {Kotochigova}\ and\ \citenamefont
  {Petrov}(2011)}]{Kotochigova2011}%
  \BibitemOpen
  \bibfield  {author} {\bibinfo {author} {\bibfnamefont {S.}~\bibnamefont
  {Kotochigova}}\ and\ \bibinfo {author} {\bibfnamefont {A.}~\bibnamefont
  {Petrov}},\ }\href {\doibase 10.1039/c1cp21175g} {\bibfield  {journal}
  {\bibinfo  {journal} {Physical chemistry chemical physics : PCCP}\ }\textbf
  {\bibinfo {volume} {13}},\ \bibinfo {pages} {19165} (\bibinfo {year}
  {2011})}\BibitemShut {NoStop}%
\bibitem [{\citenamefont {de~Marchi}\ \emph {et~al.}(1984)\citenamefont
  {de~Marchi}, \citenamefont {Rovera},\ and\ \citenamefont
  {Premoli}}]{Marchi1984}%
  \BibitemOpen
  \bibfield  {author} {\bibinfo {author} {\bibfnamefont {A.}~\bibnamefont
  {de~Marchi}}, \bibinfo {author} {\bibfnamefont {G.~D.}\ \bibnamefont
  {Rovera}}, \ and\ \bibinfo {author} {\bibfnamefont {A.}~\bibnamefont
  {Premoli}},\ }\href {\doibase 10.1088/0026-1394/20/2/002} {\bibfield
  {journal} {\bibinfo  {journal} {Metrologia}\ }\textbf {\bibinfo {volume}
  {20}},\ \bibinfo {pages} {37} (\bibinfo {year} {1984})}\BibitemShut {NoStop}%
\bibitem [{\citenamefont {Beyer}\ \emph {et~al.}(2015)\citenamefont {Beyer},
  \citenamefont {Maisenbacher}, \citenamefont {Khabarova}, \citenamefont
  {Matveev}, \citenamefont {Pohl}, \citenamefont {Udem}, \citenamefont
  {H{\"{a}}nsch},\ and\ \citenamefont {Kolachevsky}}]{Beyer2015}%
  \BibitemOpen
  \bibfield  {author} {\bibinfo {author} {\bibfnamefont {A.}~\bibnamefont
  {Beyer}}, \bibinfo {author} {\bibfnamefont {L.}~\bibnamefont {Maisenbacher}},
  \bibinfo {author} {\bibfnamefont {K.}~\bibnamefont {Khabarova}}, \bibinfo
  {author} {\bibfnamefont {A.}~\bibnamefont {Matveev}}, \bibinfo {author}
  {\bibfnamefont {R.}~\bibnamefont {Pohl}}, \bibinfo {author} {\bibfnamefont
  {T.}~\bibnamefont {Udem}}, \bibinfo {author} {\bibfnamefont {T.~W.}\
  \bibnamefont {H{\"{a}}nsch}}, \ and\ \bibinfo {author} {\bibfnamefont
  {N.}~\bibnamefont {Kolachevsky}},\ }\href {\doibase
  10.1088/0031-8949/2015/T165/014030} {\bibfield  {journal} {\bibinfo
  {journal} {Physica Scripta}\ }\textbf {\bibinfo {volume} {T165}},\ \bibinfo
  {pages} {014030} (\bibinfo {year} {2015})}\BibitemShut {NoStop}%
\bibitem [{\citenamefont {Nicholson}\ \emph {et~al.}(2015)\citenamefont
  {Nicholson}, \citenamefont {Campbell}, \citenamefont {Hutson}, \citenamefont
  {Marti}, \citenamefont {Bloom}, \citenamefont {McNally}, \citenamefont
  {Zhang}, \citenamefont {Barrett}, \citenamefont {Safronova}, \citenamefont
  {Strouse}, \citenamefont {Tew},\ and\ \citenamefont {Ye}}]{Nicholson2015}%
  \BibitemOpen
  \bibfield  {author} {\bibinfo {author} {\bibfnamefont {T.~L.}\ \bibnamefont
  {Nicholson}}, \bibinfo {author} {\bibfnamefont {S.~L.}\ \bibnamefont
  {Campbell}}, \bibinfo {author} {\bibfnamefont {R.~B.}\ \bibnamefont
  {Hutson}}, \bibinfo {author} {\bibfnamefont {G.~E.}\ \bibnamefont {Marti}},
  \bibinfo {author} {\bibfnamefont {B.~J.}\ \bibnamefont {Bloom}}, \bibinfo
  {author} {\bibfnamefont {R.~L.}\ \bibnamefont {McNally}}, \bibinfo {author}
  {\bibfnamefont {W.}~\bibnamefont {Zhang}}, \bibinfo {author} {\bibfnamefont
  {M.~D.}\ \bibnamefont {Barrett}}, \bibinfo {author} {\bibfnamefont {M.~S.}\
  \bibnamefont {Safronova}}, \bibinfo {author} {\bibfnamefont {G.~F.}\
  \bibnamefont {Strouse}}, \bibinfo {author} {\bibfnamefont {W.~L.}\
  \bibnamefont {Tew}}, \ and\ \bibinfo {author} {\bibfnamefont
  {J.}~\bibnamefont {Ye}},\ }\href {\doibase 10.1038/ncomms7896} {\bibfield
  {journal} {\bibinfo  {journal} {Nature communications}\ }\textbf {\bibinfo
  {volume} {6}},\ \bibinfo {pages} {6896} (\bibinfo {year} {2015})}\BibitemShut
  {NoStop}%
\bibitem [{\citenamefont {Ishikawa}\ \emph {et~al.}(1997)\citenamefont
  {Ishikawa}, \citenamefont {Hatakeyama}, \citenamefont {Gosyono-o},
  \citenamefont {Wada}, \citenamefont {Takahashi},\ and\ \citenamefont
  {Yabuzaki}}]{ishikawa1997}%
  \BibitemOpen
  \bibfield  {author} {\bibinfo {author} {\bibfnamefont {K.}~\bibnamefont
  {Ishikawa}}, \bibinfo {author} {\bibfnamefont {A.}~\bibnamefont
  {Hatakeyama}}, \bibinfo {author} {\bibfnamefont {K.}~\bibnamefont
  {Gosyono-o}}, \bibinfo {author} {\bibfnamefont {S.}~\bibnamefont {Wada}},
  \bibinfo {author} {\bibfnamefont {Y.}~\bibnamefont {Takahashi}}, \ and\
  \bibinfo {author} {\bibfnamefont {T.}~\bibnamefont {Yabuzaki}},\ }\href
  {\doibase 10.1103/PhysRevB.56.780} {\bibfield  {journal} {\bibinfo  {journal}
  {Physical Review B}\ }\textbf {\bibinfo {volume} {56}},\ \bibinfo {pages}
  {780} (\bibinfo {year} {1997})}\BibitemShut {NoStop}%
\bibitem [{\citenamefont {Chin}\ \emph {et~al.}(2010)\citenamefont {Chin},
  \citenamefont {Grimm}, \citenamefont {Julienne},\ and\ \citenamefont
  {Tiesinga}}]{Chin2010}%
  \BibitemOpen
  \bibfield  {author} {\bibinfo {author} {\bibfnamefont {C.}~\bibnamefont
  {Chin}}, \bibinfo {author} {\bibfnamefont {R.}~\bibnamefont {Grimm}},
  \bibinfo {author} {\bibfnamefont {P.}~\bibnamefont {Julienne}}, \ and\
  \bibinfo {author} {\bibfnamefont {E.}~\bibnamefont {Tiesinga}},\ }\href
  {\doibase 10.1103/RevModPhys.82.1225} {\bibfield  {journal} {\bibinfo
  {journal} {Reviews of Modern Physics}\ }\textbf {\bibinfo {volume} {82}},\
  \bibinfo {pages} {1225} (\bibinfo {year} {2010})}\BibitemShut {NoStop}%
\bibitem [{\citenamefont {Kolachevsky}\ \emph {et~al.}(2007)\citenamefont
  {Kolachevsky}, \citenamefont {Akimov}, \citenamefont {Tolstikhina},
  \citenamefont {Chebakov}, \citenamefont {Sokolov}, \citenamefont {Rodionov},
  \citenamefont {Kanorski},\ and\ \citenamefont {Sorokin}}]{Kolachevsky2007}%
  \BibitemOpen
  \bibfield  {author} {\bibinfo {author} {\bibfnamefont {N.}~\bibnamefont
  {Kolachevsky}}, \bibinfo {author} {\bibfnamefont {A.}~\bibnamefont {Akimov}},
  \bibinfo {author} {\bibfnamefont {I.}~\bibnamefont {Tolstikhina}}, \bibinfo
  {author} {\bibfnamefont {K.}~\bibnamefont {Chebakov}}, \bibinfo {author}
  {\bibfnamefont {A.}~\bibnamefont {Sokolov}}, \bibinfo {author} {\bibfnamefont
  {P.}~\bibnamefont {Rodionov}}, \bibinfo {author} {\bibfnamefont
  {S.}~\bibnamefont {Kanorski}}, \ and\ \bibinfo {author} {\bibfnamefont
  {V.}~\bibnamefont {Sorokin}},\ }\href {\doibase 10.1007/s00340-007-2835-z}
  {\bibfield  {journal} {\bibinfo  {journal} {Applied Physics B}\ }\textbf
  {\bibinfo {volume} {89}},\ \bibinfo {pages} {589} (\bibinfo {year}
  {2007})}\BibitemShut {NoStop}%
\bibitem [{\citenamefont {Friebel}\ \emph {et~al.}(1998)\citenamefont
  {Friebel}, \citenamefont {{D Andrea}}, \citenamefont {Walz}, \citenamefont
  {Weitz},\ and\ \citenamefont {Haensch}}]{Friebel1998}%
  \BibitemOpen
  \bibfield  {author} {\bibinfo {author} {\bibfnamefont {S.}~\bibnamefont
  {Friebel}}, \bibinfo {author} {\bibfnamefont {C.}~\bibnamefont {{D Andrea}}},
  \bibinfo {author} {\bibfnamefont {J.}~\bibnamefont {Walz}}, \bibinfo {author}
  {\bibfnamefont {M.}~\bibnamefont {Weitz}}, \ and\ \bibinfo {author}
  {\bibfnamefont {T.~W.}\ \bibnamefont {Haensch}},\ }\href {\doibase
  10.1103/PhysRevA.57.R20} {\bibfield  {journal} {\bibinfo  {journal} {Physical
  Review A}\ }\textbf {\bibinfo {volume} {57}},\ \bibinfo {pages} {R20}
  (\bibinfo {year} {1998})}\BibitemShut {NoStop}%
\bibitem [{\citenamefont {Landau}\ and\ \citenamefont
  {Lifshits}(1976)}]{Landau:Mechanics}%
  \BibitemOpen
  \bibfield  {author} {\bibinfo {author} {\bibfnamefont {L.~D.}\ \bibnamefont
  {Landau}}\ and\ \bibinfo {author} {\bibfnamefont {E.~M.}\ \bibnamefont
  {Lifshits}},\ }\href
  {https://books.google.com/books?id=e-xASAehg1sC&printsec=frontcover&hl=ru&source=gbs_ge_summary_r&cad=0#v=onepage&q&f=false}
  {\emph {\bibinfo {title} {{Mechnics}}}},\ \bibinfo {edition} {3rd}\ ed.\
  (\bibinfo  {publisher} {Pergamon press},\ \bibinfo {year} {1976})\BibitemShut
  {NoStop}%
\bibitem [{\citenamefont {Martin}\ \emph {et~al.}(1978)\citenamefont {Martin},
  \citenamefont {Zalubas},\ and\ \citenamefont {Hagan}}]{TmEnergyLevels}%
  \BibitemOpen
  \bibfield  {author} {\bibinfo {author} {\bibfnamefont {W.~C.}\ \bibnamefont
  {Martin}}, \bibinfo {author} {\bibfnamefont {R.}~\bibnamefont {Zalubas}}, \
  and\ \bibinfo {author} {\bibfnamefont {L.}~\bibnamefont {Hagan}},\
  }\href@noop {} {\emph {\bibinfo {title} {{Atomic Energy Levels: The
  Rare-Earth Elements}}}}\ (\bibinfo  {publisher} {Nat. Bur. Stand., U.S.},\
  \bibinfo {year} {1978})\BibitemShut {NoStop}%
\bibitem [{\citenamefont {Chernov}\ \emph {et~al.}(2005)\citenamefont
  {Chernov}, \citenamefont {Dorofeev}, \citenamefont {Kretinin},\ and\
  \citenamefont {Zon}}]{Chernov2005}%
  \BibitemOpen
  \bibfield  {author} {\bibinfo {author} {\bibfnamefont {V.~E.}\ \bibnamefont
  {Chernov}}, \bibinfo {author} {\bibfnamefont {D.~L.}\ \bibnamefont
  {Dorofeev}}, \bibinfo {author} {\bibfnamefont {I.~Y.}\ \bibnamefont
  {Kretinin}}, \ and\ \bibinfo {author} {\bibfnamefont {B.~A.}\ \bibnamefont
  {Zon}},\ }\href {\doibase 10.1088/0953-4075/38/13/020} {\bibfield  {journal}
  {\bibinfo  {journal} {Journal of Physics B}\ }\textbf {\bibinfo {volume}
  {38}},\ \bibinfo {pages} {2289} (\bibinfo {year} {2005})}\BibitemShut
  {NoStop}%
\end{thebibliography}%

\end{document}